\documentclass[aps,showpacs,a4paper,floatfix,twocolumn,prc]{revtex4-1}

\usepackage[colorlinks,citecolor=blue,linkcolor=blue,anchorcolor=blue,filecolor=blue,
urlcolor=blue]{hyperref}

\usepackage{graphicx,color,xcolor}
\usepackage{amsfonts,amsmath}
\usepackage{amssymb,ulem}
\usepackage{dcolumn}
\usepackage{bm,lipsum}
\usepackage[utf8]{inputenc}
\usepackage{subeqnarray}
\usepackage{array}
\usepackage{epsfig,hyperref}
\usepackage{morefloats}
\usepackage{relsize}
\usepackage{url}

\usepackage{comment}

\begin{document}

\preprint{ }

\title{
The baryon coupling scheme in an unified  SU(3) and SU(6) symmetry formalism}

\author{Luiz L. Lopes}
\email{llopes@cefetmg.br}

\affiliation{%
 Centro Federal de Educa\c{c}\~ao Tecnol\'ogica de Minas Gerais Campus VIII, Varginha/MG, CEP 37.022-560, Brazil
}%

\author{Kauan D. Marquez}
\affiliation{Departamento de F\'{\i}sica - CFM, Universidade Federal de Santa Catarina,  Florian\'opolis/SC, CEP 88.040-900, Brazil}

\author{Débora P. Menezes}
\affiliation{Departamento de F\'{\i}sica - CFM, Universidade Federal de Santa Catarina,  Florian\'opolis/SC, CEP 88.040-900, Brazil}

\date{\today}

\begin{abstract}
 We calculate the baryon-meson coupling constants for the spin-1/2 baryonic octet and spin-3/2 decuplet in a unified approach relying on symmetry arguments such as the fact that the Yukawa couplings, present in the Lagrangian density of the Walecka-type models, must be an invariant under SU(3) and SU(6) group transformations. The coupling constants of the baryon with the scalar $\sigma$ meson are fixed to reproduce the known potential depths for the hyperons and $\Delta$ resonances, in  an approach that can be extended to all particles. We then apply the calculated coupling constants to study neutron star matter with hyperons and deltas admixed to its composition. We conclude that the $\Delta^-$ is by far the most important exotic particle that can be present in the neutron star interior. It is always present, independent of the chosen parameterization, and might appear in almost every known neutron star, once its onset happens at very low density. Yet, its presence affects the astrophysical properties of the canonical 1.4 M$_\odot$ star, and, in some cases, it can even contribute to an increase in the maximum mass reached.
\end{abstract}


\maketitle

\section{Introduction}

{Knowing the neutron star inner composition is a central piece of the research program on these objects. The first theoretical descriptions of neutron stars (NS) by Landau, Baade and Zwicky in the 1930s considered protons and electrons alongside with the neutrons,  in order to acknowledge the stability of the nuclear matter under $\beta$ decay and the charge neutrality condition \cite{Landau,incluir}.

Different non-nucleonic degrees of freedom are considered in the literature when describing neutron star matter. The inclusion of the entire spin-1/2 baryon octet (i.e., nucleons and hyperons) is almost standard to the theoretical description of such objects \cite{Glen}.}
These studies gave rise to the very much discussed hyperon puzzle: on one hand, their inclusion is a direct follow up of energetical considerations but, on the other, they soften the equation of state (EOS) resulting in a decrease of the maximum stellar mass attained \cite{hyperonpuzzle, Universe}, which is not desirable since the first really massive NS was detected with a robust relativistic Shapiro-delay method in the 2010s \cite{Antoniadis}. From the theoretical side, a clear caveat is that all calculations depend on the hyperon-meson couplings and some of them are poorly defined. One possible solution to fix these quantities in a less arbitrary way is to use group theory, as previoulsly done in \cite{Weiss2, Lopes2013, lopesnpa,Tsu}.

From the experimental and observational side, along the COVID-19 pandemic (2020-2021), the NICER telescope kept our interests very much alive with some new results for the NS radii. Surprisingly a massive NS and a canonical one (with a mass of the order of 1.4 M$_\odot$) bear quite similar radii \cite{NICER3,Miller}. Meanwhile, the LIGO and Virgo collaboration remained fully operational, detecting more gravitational waves and providing more information on tidal polarizabilities, some of which will be used to analyse our results along the present work \cite{tidal1}.

Even being of huge value to the improvement of the theoretical description of compact stars, these results are far from giving a conclusive picture of the internal composition of such objects. Thus, it is still interesting to explore the influence of other exotic particles in  the neutron star EOS. An obvious next step is the inclusion of the baryons of the spin-3/2 decuplet. In the latest years, many papers addressed this fundamental question, specially considering the $\Delta$ resonances \cite{SergioDuarte, Sedrakian, nossoVeronica, nossoConstanca}. As it is with the hyperons, the couplings of the spin 3/2 particles with the mesons are very important but remain largely unknown. A very broad study about the delta-meson coupling effects on the NS description was performed in Ref. \cite{nossoConstanca}, but the authors let the delta couplings vary freely within a given range.

In the present work, we fix the couplings of the octet and of the decuplet baryons with the mesons through an unified SU(3) and SU(6) group symmetry \cite{Lopes2013,lopesnpa,Lopes2020a,Weiss1,Weiss2,Dover1984,Tsu,MM1979,Stancu,Rijken,Lipkin,Swart1963}. This is a good recipe to get rid of the huge arbitrariness previously used, while all the baryonic potentials are not obtained experimentally. Once the Clebsch-Gordan coefficients are used to calculate all the couplings, just one free parameter is left to be freely varied. The resulting EOS are then obtained and the corresponding mass-radius diagrams are plotted and discussed. The adiabatic index and the tidal polarizability are also checked. For a better understanding of the effects of the inclusion of the hyperons and the $\Delta$ particles in our calculations, we present our results step-by-step: we first consider only nucleons and hyperons, as done in \cite{lopesnpa,Lopes2013,Weiss1,Weiss2,Tsu}, then we consider only nucleons and $\Delta$'s and finally we include the whole octet and the $\Delta$ particles in our calculations. It is worth mentioning that the prescription for the calculation of the decuplet-meson couplings presented in this work can be applied to a diversity of calculations in future works. 

\section{The coupling constants with the vector mesons}

Let's assume that the Yukawa couplings of the quantum hadrodynamics (QHD) Lagrangian, of the type
\begin{equation}
\mathcal{L}_{\rm Yukawa} = -(g_{BBM})(\bar{\psi}_B\psi_B)M, \label{Y1}
\end{equation}
where $\psi_B$ is the Dirac field of the baryon $B$, and $M$ is an arbitrary meson ~\cite{Swart1963}, are invariant under the flavor SU(3) symmetry group and that the more restrictive hybrid  $SU(6) \supset SU(3) \otimes SU(2)$ group is only partially broken. The Yukawa coupling constant of the baryon $B$ with the meson $M$, given by  $g_{BBM}$, can be written in terms of only one free parameter $\alpha_v$ (see the appendix for a detailed discussion).
The relative strength of the coupling between the mesons and the exotic baryons will be given by
\begin{align}
 \frac{g_{\Lambda\Lambda\omega}}{g_{NN\omega}} ={}& \frac{4 + 2\alpha_v}{5 + 4 \alpha_v},\\
 \frac{g_{\Sigma\Sigma\omega}}{g_{NN\omega}} ={}& \frac{8 - 2\alpha_v}{5 + 4 \alpha_v}\\
  \frac{g_{\Xi\Xi\omega}}{g_{NN\omega}} ={}& \frac{5 - 2\alpha_v}{5 + 4 \alpha_v}, \label{e1}
\end{align}
for the $\omega$ meson,
\begin{align}
 \frac{g_{\Lambda\Lambda\phi}}{g_{NN\omega}}={}& -\sqrt{2} \left(\frac{ 5-2\alpha_v}{5 + 4 \alpha_v} \right),\\
 \frac{g_{\Sigma\Sigma\phi}}{g_{NN\omega}} ={}&- \sqrt{2} \left(\frac{ 1+ 2\alpha_v }{5 + 4 \alpha_v} \right),\\
  \frac{g_{\Xi\Xi\phi}}{g_{N,\omega}} ={}&- \sqrt{2} \left(\frac{ 4+ 2\alpha_v}{5 + 4 \alpha_v} \right),\label{e2}
\end{align}
with $\frac{g_{NN\phi}}{g_{NN\omega}} = 0$ for the $\phi$ meson, and
\begin{align}
 \frac{g_{\Lambda\Lambda\rho}}{g_{NN\rho}} ={}& 0,\\
 \frac{g_{\Sigma\Sigma\rho}}{g_{N,\rho}} ={}& 2\alpha_v,\\
  \frac{g_{\Xi\Xi\rho}}{g_{NN\rho}} ={}& -(1 - 2\alpha_v),\label{e3}
\end{align}
for the $\rho$ meson.

For the baryon decuplet we can also write the relative strength of the baryon-meson coupling constants in terms of the same free parameter $\alpha_v$, if we assume that the Sakurai's theory of the strong interaction (see R~\cite{Sakurai}) holds even between inter-multiplet states, as discussed in the appendix. Then, we have
\begin{align}
 \frac{g_{\Delta^*\Delta^*\omega}}{g_{NN\omega}} = \frac{g_{\Delta\Delta\omega}}{g_{NN\omega}} ={}& \frac{9}{5 + 4 \alpha_v},\\
 \frac{g_{\Sigma^*\Sigma^*\omega}}{g_{NN\omega}} ={}& \frac{6}{5 + 4 \alpha_v}, \\ 
  \frac{g_{\Xi^*\Xi^*\omega}}{g_{NN\omega}} ={}&\frac{3}{5 + 4 \alpha_v},\\\frac{g_{\Omega\Omega\omega}}{g_{NN\omega}} ={}& 0,\label{e4}
\end{align}
for the $\omega$ meson,
\begin{align}
 \frac{g_{\Sigma^*\Sigma^*\phi}}{g_{NN\omega}} ={}& \frac{-3\sqrt{2}}{5 + 4 \alpha_v},\\ 
  \frac{g_{\Xi^*\Xi^*\phi}}{g_{NN\omega}} ={}& \frac{-6\sqrt{2}}{5 + 4 \alpha_v},\\ \frac{g_{\Omega\Omega\phi}}{g_{NN\omega}} ={}& \frac{-9\sqrt{2}}{5 + 4 \alpha_v}, \label{e5}
\end{align}
with $ \frac{g_{\Delta^*\Delta^*\phi}}{g_{NN\omega}} = \frac{g_{\Delta\Delta\phi}}{g_{NN\omega}} = 0$ for the $\phi$ meson, and
\begin{align}
 \frac{g_{\Delta^*\Delta^*\rho}}{g_{NN\rho}} ={}& 3,\\  \frac{g_{\Delta\Delta\rho}}{g_{NN\rho}} ={}& 1,\\
 \frac{g_{\Sigma^*\Sigma^*\rho}}{g_{N,\rho}} ={}& 2,\\
 \frac{g_{\Xi^*\Xi^*\rho}}{g_{NN\rho}} ={}& 1, \\
 \frac{g_{\Omega\Omega\rho}}{g_{NN\rho}} ={} &0, \label{e6}
\end{align}
\noindent for the $\rho$ meson. As can be seen, no free parameter is present for the coupling with the $\rho$ meson.
Notice that 
we have combined the four $\Delta$'s into two isospin multiplets ($\Delta=\{\Delta^0, \Delta^{+}\}$ and $\Delta^\ast=\{\Delta^-, \Delta^{++}\}$). 

\section{The coupling constant with the scalar meson}

To keep the Yukawa Lagrangian density invariant under the SU(3) flavor symmetry we need the knowledge not only of the baryon eigenstates, but also of the meson ones. 
Unfortunately, unlike the vector mesons, the nature and the proper existence of the scalar mesons are still foggy and uncertain.
Nevertheless, 
despite all the beauty of the symmetry group theory, a physical theory must ultimately reproduce the results coming from the laboratory. Therefore, the relative strength of the coupling constants can be determined by the potential depths at the saturation point. The potential depth of the baryon $B$ is defined as
\begin{equation}
U_B(n_0) =  g_{BB\omega}\omega_0 - g_{BB\sigma}\sigma_0 , \label{e7}
\end{equation}
where the $\omega_0$ and the $\sigma_0$ are the RMF values of the $\omega$ and $\sigma$ fields, respectively \cite{Weiss1,lopesnpa}.
The $\Lambda$ potential depth is well known, $U_\Lambda$ = -28 MeV~\cite{Glen}, but the $U_\Sigma$ and $U_\Xi$ are known with a lesser degree of precision, although we believe that the $\Sigma$ potential is moderately repulsive, while the $\Xi$ potential is weakly attractive~\cite{Schaffner2000}. 
In this work, we use $U_\Sigma = +30$ MeV and $U_\Xi = - 4$ MeV, values that were recently favored by lattice QCD
calculations~\cite{LQCD}.

In the case of the baryon decuplet, the situation is far worse. The potential depths for the $\Delta$ baryons are
poorly constrained  and, as far as we know, there is no information about the potential depth of the other species of the baryon decuplet. Therefore, we present no value for the scalar meson coupling constant for these baryons.
As extensively discussed in Ref.~\cite{Delta2017}, different models and techniques
point to a potential depth $- 150$ MeV $< U_\Delta < - 50 $ MeV. A more recent study~\cite{Delta2020} suggests that  $U_\Delta~\approx~1.5~U_N$. Here, we assume the latter value.

Unfortunately, the values of the $\omega$ and $\sigma$ fields in Eq.~(\ref{e7}) are not model independent, implying that the $g_{BB\sigma}$ will not be independently determined either. Thus, to determine the couplings with the scalar meson while still satisfying nuclear bulk properties, we need to employ a more comprehensive Lagrangian density, which involves the kinetic terms of baryons and mesons in addition to the Yuakawa couplings from Eq.~\eqref{Y1}. In general, non-linear terms and their couplings are also necessary. 
In this work we use an enhanced version of the L3$\omega\rho$  QHD parameterization presented in Ref.~\cite{lopes2022}, which virtually satisfies every single constraint of the symmetric nuclear matter at the saturation point. Within this model, we have $U_\Delta = 1.5~U_N = -98$ MeV.  

A different kind of constraint was presented in Ref.~\cite{Delta1989}, and it is related to the difference
between the relative strength of the scalar channel 
and the vector one, that must obey the relation
\begin{equation}
0 ~ \leq ~\left(\frac{g_{\Delta\Delta\sigma}}{g_{NN\sigma}} - \frac{g_{\Delta\Delta\omega}}{g_{NN\omega}} \right) ~ \leq ~ 0.2 . \label{e1989}
\end{equation}
As we show next, for our chosen parameterization and potential depth, such constraint is always satisfied.

The non-linear Walecka-type QHD Lagrangian employed here reads
\begin{widetext}
\begin{align}
\mathcal{L}_{\rm QHD} ={}& \sum_B \bar{\psi}_B\left[\gamma^\mu\left(\mbox{i}\partial_\mu  - g_{BB\omega}\omega_\mu -  g_{BB\phi}\phi_\mu  -  \frac{g_{BB\rho}}{2}\vec{\tau} \cdot \vec{\rho}_\mu\right)- (M_B - g_{BB\sigma}\sigma)\right]\psi_B      \nonumber   \\
& + \frac{1}{2}(\partial_\mu \sigma \partial^\mu \sigma - m_s^2\sigma^2) - \frac{\kappa}{3} M_N(g_{\sigma} \sigma)^3 - \frac{\lambda}{4}(g_{\sigma}\sigma)^4 - \frac{1}{4}\Omega^{\mu \nu}\Omega_{\mu \nu} + \frac{1}{2} m_v^2 \omega_\mu \omega^\mu\nonumber\\
&-  \frac{1}{4}{\bf{P}^{\mu \nu} \cdot {P}_{\mu \nu} }+ \frac{1}{2} m_\rho^2 \vec{\rho}_\mu \cdot \vec{\rho}^{ \; \mu}  - \frac{1}{4}\Phi^{\mu\nu}\Phi_{\mu\nu}  + \frac{1}{2}m_\phi^2\phi_\mu\phi^\mu  + \Lambda_{\omega\rho}(g_{\rho}^2 \vec{\rho^\mu} \cdot \vec{\rho_\mu}) (g_{\omega}^2 \omega^\mu \omega_\mu),
\label{e8} 
\end{align}
\end{widetext}
where the $\psi_B$  represent both the baryonic  Dirac field of the baryon octet and of the decuplet, with mass $M_B$. We are aware that from a rigorous point of view, the members of the baryon decuplet should be described by the Rarita-Schwinger Lagrangian density. Nonetheless, as shown in Ref.~\cite{Paoli2},  the resulting equation of motion can be written compactly as a Dirac equation with the same energy eigenvalue.  The $i=\sigma, \omega_\mu,\phi,\vec{\rho}_\mu$ are the mesonic fields with mass $m_i$. The $g_{BBi}$ are the Yukawa coupling constants that simulate the strong interaction between  the baryon $B$ and the  meson $i$, and $\vec{\tau}$ are the Pauli matrices.
 The antisymmetric mesonic field strength tensors are given by their usual expressions as presented in~\cite{Glen}. The $\Lambda_{\omega\rho}$ term represents a non-linear  $\omega$-$\rho$ coupling between the vector mesons as, e.g., the one present in the IUFSU model~\cite{IUFSU}. 
 The parameters utilized in this work and the predictions to nuclear matter properties are presented in Tab.~\ref{TL0}. The meson masses are $m_\omega$ = 783 MeV, $m_\rho$ = 770 MeV, $m_\phi$ = 1020 MeV, $m_\sigma$ = 512 MeV

 The main experimental constraints are related  to the nuclear matter saturation point. The acceptable values for the saturation density itself ($n_0$) the effective nucleon mass ($M_N^*/M_N$),  the binding energy per nucleon ($B/A$) and the compressibility ($K$) are taken from two extensive review articles, see Refs. ~\cite{Dutra2014,Micaela2017}.
 The values of the  symmetry energy  ($S_0$) and its slope ($L$)  were strongly constrained  combining astrophysical data with nuclear properties measured in the PREX-II experiment, together with chiral effective field theory, in Ref.~\cite{cPrex}. We use {31.2 MeV $<~S_0<$ 35 MeV}, and {38 MeV $<~L~<$ 67 MeV} as bounds to these quantities.
 
 \begin{table}[!t]
     \centering
     \begin{tabular}{c|c}
     \hline
     \multicolumn{2}{c}{Enhanced L3$\omega\rho$}
     \\ \hline
    $(g_{NN\sigma}/m_s)^2$ & 12.108 fm$^2$ \\
    $(g_{NN\omega}/m_v)^2$ & 7.132  fm$^2$ \\
    $(g_{NN\rho}/m_\rho)^2$ & 5.85  fm$^2$ \\
    $\kappa$ & 0.04138 \\
    $\lambda$ &  $-0.0390$ \\
    $\Lambda_{\omega\rho}$ &  0.0283 \\ \hline
     \end{tabular}
     
\begin{tabular}{cc}
    { }&{ } \\
\end{tabular}

\begin{tabular}{c|cc}
\hline 
Quantity & Constraint & This model\\\hline
$n_0$ ($fm^{-3}$) & 0.148--0.170 & 0.156 \\
   $M^{*}/M$ & 0.60--0.80 & 0.69  \\

  $K$ (MeV)& 220--260                                          &  256  \\

 $S_0$ (MeV) & 31.2--35.0 &  32.1  \\
$L$ (MeV) & 38--67 & 66\\

 $B/A$ (MeV) & 15.8--16.5  & 16.2  \\ 
 $S(2n_0)$ (MeV) & 38.0--64.0  & 49.8  \\ 
 $p(2n_0)$ (MeV/fm$^3$) & 11.2--38.7  & 16.4  \\ 
\hline
\end{tabular}
 
\caption{ Enhanced version of the L3$\omega\rho$~\cite{lopes2022} parameterization (top) and its predictions to nuclear matter (bottom). The phenomenological constraints are taken from Ref.~\cite{Dutra2014,Micaela2017,cPrex,s3,tidal1}. } 
\label{TL0}
\end{table}

Another two recent constraints are set beyond the saturation point. They are the symmetry energy at twice the saturation density, $S(2n_0)$, and the pressure of the symmetric matter at the same point, $p(2n_0)$.
The authors in Ref.~\cite{s3} bound $S(2n_0)$ in the range between 51 $\pm$ 13MeV at 68$\%$ conﬁdence level, while Ref.~\cite{tidal1} ﬁxed $p(2n_0)$ between 11.2 MeV$\cdot$fm$^{-3}$ and 38.7 Mev$\cdot$fm$^{-3}$ at
the 90$\%$ level. Our enhanced
L3$\omega\rho$ parametrization agrees with all constraint presented.

 
 
 
 To solve the equations of motion, we use the mean field approximation, where the meson fields are replaced by their expectation values. Applying the Euler-Lagrange formalism, and using the quantization rules ($E = \partial^0$ , $k = i\partial^j$) we easily obtain the eigenvalue for the energy
\begin{equation}
 E_B = \sqrt{k^2 + M_B^{*2}} + g_{B\omega}\omega_0 + g_{B\phi}\phi_0 + \frac{\tau_{3B}}{2}g_{B\rho}\rho_0 , \label{EL9}
\end{equation}
where $M^{*}_B~\equiv~M_B - g_{B\sigma}\sigma_0$ is the effective baryon mass and $\tau_{3B}$ is the isospin projection of the baryon.
A non-interacting lepton gas is also included.

Applying Fermi-Dirac
statistics to baryons and leptons, we can write the total energy density as
\begin{align}
 \epsilon ={}& \sum_B \frac{\gamma}{2\pi^2}\int_0^{k_{Bf}} dk k^2 \sqrt{k^2 + M_B^{*2}} \nonumber \\
 &+\frac{1}{2}m_\sigma^2\sigma_0^2 + \frac{1}{2}m_\omega^2\omega_0^2 + \frac{1}{2}m_\phi^2\phi_0^2 + \frac{1}{2}m_\rho^2\rho_0^2   \nonumber \\ & + \frac{\kappa }{3}M_N(g_{\sigma} \sigma)^3 +  \frac{\lambda}{4} (g_{\sigma}\sigma)^4 +3 \Lambda_v\omega_0^2\rho_0^2 
\nonumber\\& + \sum_l \frac{1}{\pi^2}\int_0^{k_{lf}} dk k^2 \sqrt{k^2 + m_l^{2}}  \label{EL11}
\end{align}
where $\Lambda_v~\equiv~\Lambda_{\omega\rho}g_{N\omega}^2g_{N\rho}^2$; $\gamma$ is the degeneracy factor $\gamma = (2S+1$), and assumes $\gamma=2$ to the members of the baryon octet and $\gamma=4$ for the members of the baryon decuplet. 

The pressure is easily obtained by thermodynamic relations: $p =\sum_f \mu_f n_f - \epsilon$, where the sum runs over all the fermions and $\mu_f$ is the corresponding chemical potential. In order to ensure the chemical equilibrium condition, the relation
\begin{equation}
    \mu_B=\mu_n-q_B\mu_e
\end{equation}
must hold, where $\mu_n$ and $\mu_e$ are the chemical potentials of the nucleon and electron, respectively, $q_B$ is the electric charge of the baryon, and $\mu_\mu=\mu_e$.
Additional discussion about the formalism can be found in Refs.~\cite{Tsu,Lopes2013,lopesnpa} and the references therein.

To obtain  $g_{BB\sigma}$ from Eq.~(\ref{e7}), we determine the fields considering symmetric nuclear matter, i.e, containing only protons and neutrons with equal densities and no leptons. 
Fixing the potentials:  $U_\Lambda = -28$ MeV, $U_\Sigma = +30$ MeV, $U_\Xi = -4$ MeV, and $U_\Delta = -98$ MeV,  we are able to determine the whole set of coupling constants for each value of $\alpha_v$, presented in Tab~\ref{T2}. Notice that, for any value of $\alpha_v$, the constraint presented in Eq.~(\ref{e1989}) is always fulfilled.

\begin{table}[!t]
\begin{center}
\begin{tabular}{c|cccc}
&\multicolumn{4}{c}{$\alpha_v$}\\
  &   1.00&0.75&0.50&0.25   \\
\hline
 $g_{\Lambda\Lambda\omega}/g_{NN\omega}$        & 0.667 & 0.687   & 0.714 & 0.75   \\

 $g_{\Sigma\Sigma\omega}/g_{NN\omega}$         & 0.667 & 0.812  & 1.0 & 1.25   \\
 
$g_{\Xi\Xi\omega}/g_{NN\omega}$           & 0.333 & 0.437  & 0.571 & 0.75   \\
\hline
$g_{\Lambda\Lambda\phi}/g_{NN\omega}$  & -0.471 & -0.619  & -0.808 & -1.06   \\

$g_{\Sigma\Sigma\phi}/g_{NN\omega}$           & -0.471 & -0.441  & -0.404 & -0.354   \\

$g_{\Xi\Xi\phi}/g_{NN\omega}$           & -0.943 & -0.972  & -1.01 & -1.06   \\
\hline
$g_{\Lambda\Lambda\rho}/g_{NN\rho}$  &0.0&0.0&0.0&0.0\\
$g_{\Sigma\Sigma\rho}/g_{NN\rho}$           & 2.0 & 1.5  & 1.0 & 0.5   \\

$g_{\Xi\Xi\rho}/g_{NN\rho}$           & 1.0 & 0.5  & 0.0 & -0.5   \\
\hline
$g_{\Lambda\Lambda\sigma}/g_{NN\sigma}$           & 0.610 & 0.625  & 0.646 & 0.674   \\

$g_{\Sigma\Sigma\sigma}/g_{NN\sigma}$           & 0.406 & 0.518  & 0.663 & 0.855   \\

$g_{\Xi\Xi\sigma}/g_{NN\sigma}$           & 0.269 & 0.350 & 0.453 & 0.590   \\
\hline
\hline
$g_{\Delta\Delta\omega}/g_{NN\omega}$        & 1.0 & 1.125  & 1.285 & 1.5   \\

$g_{\Delta^*\Delta^*\omega}/g_{NN\omega}$        & 1.0 & 1.125   & 1.285 & 1.5 \\

 $g_{\Sigma^*\Sigma^*\omega}/g_{NN\omega}$         & 0.667 & 0.75  & 0.857 & 1.0   \\
 
$g_{\Xi^*\Xi^*\omega}/g_{NN\omega}$           & 0.333 & 0.375  & 0.428 & 0.667   \\

$g_{\Omega\Omega\omega}/g_{NN\omega}$           & 0.0 & 0.0  & 0.0 & 0.0   \\
\hline
$g_{\Sigma^*\Sigma^*\phi}/g_{NN\omega}$         & -0.471 & -0.530  & -0.606 & -0.707   \\
 
$g_{\Xi^*\Xi^*\phi}/g_{NN\omega}$           & -0.943 & -1.060 & -1.212 & -1.414   \\

$g_{\Omega\Omega\phi}/g_{NN\omega}$           & -1.414 & -1.590  & -1.818 & -2.212   \\
\hline
$g_{\Delta\Delta\rho}/g_{NN\rho}$        & 1.0 & 1.0  & 1.0 & 1.0   \\

$g_{\Delta^*\Delta^*\rho}/g_{NN\rho}$        & 3.00 & 3.0   & 3.0 & 3.0 \\

 $g_{\Sigma^*\Sigma^*\rho}/g_{NN\rho}$         & 2.00 & 2.0  & 2.0 & 2.0  \\
 
$g_{\Xi^*\Xi^*\rho}/g_{NN\rho}$           & 1.0 & 1.0  & 1.0 & 1.0   \\
$g_{\Omega\Omega\rho}/g_{NN\rho}$           & 0.0 & 0.0  & 0.0 & 0.0   \\
\hline
$g_{\Delta\Delta\sigma}/g_{NN\sigma}$        & 1.110 & 1.208  & 1.331 & 1.5  \\

$g_{\Delta^*\Delta^*\sigma}/g_{NN\sigma}$        & 1.110 & 1.208   & 1.331 & 1.5 \\

 $g_{\Sigma^*\Sigma^*\sigma}/g_{NN\sigma}$         & ? & ?  & ? & ?   \\
 
$g_{\Xi^*\Xi^*\sigma}/g_{NN\sigma}$           & ? & ?  & ? & ?   \\

$g_{\Omega\Omega\sigma}/g_{NN\sigma}$           & ? & ?  & ? & ?   \\
\hline

\end{tabular}
 \caption{Baryon-meson coupling constants for different values of $\alpha_v$ to reproduce $U_\Lambda = -28$ MeV, $U_\Sigma = +30$ MeV, $U_\Xi = -4$ MeV, and $U_\Delta = -98$ MeV. The potentials depth for the $\Sigma^{*}$, $\Xi^{*}$ and $\Omega$ are still unknown, so it is not possible to determine their scalar couplings with the current  knowledge.}\label{T2}
 \end{center}
 \end{table}

\section{Hyperonic neutron stars}

We start by describing the features of hyperonic neutron stars and their constituents. The results presented in this section are not new, as they have already been discussed in the past~\cite{lopesnpa,Lopes2013,Weiss2,Tsu}, but they are important to allow us a direct comparison between the effect of hyperons and $\Delta$ as as well as with $\Delta$-admixed hyperonic nuclear matter.

\begin{figure*}[!t]
\centering
\includegraphics[width=.3\textwidth, angle=270]{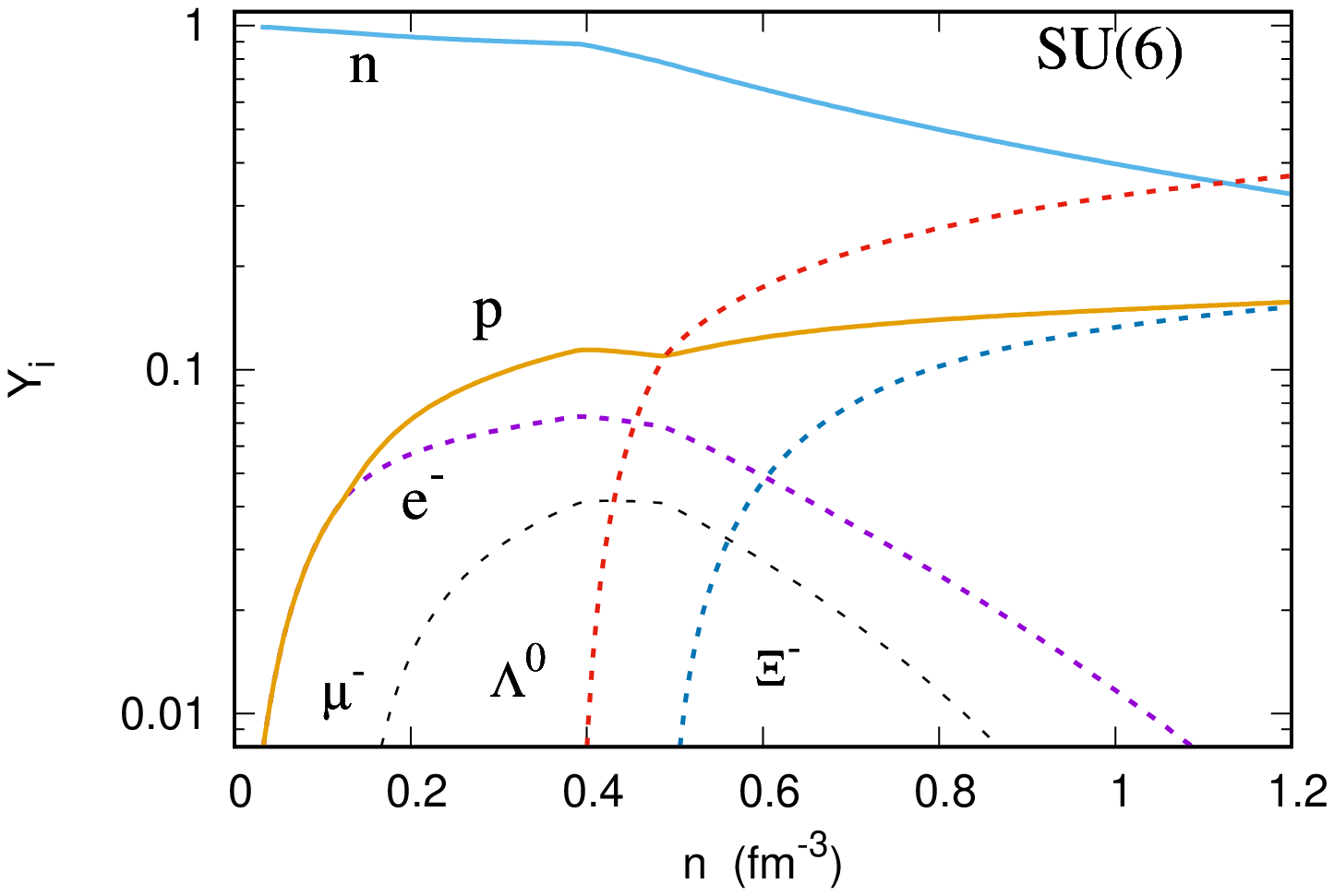} 
\includegraphics[width=.3\textwidth, angle=270]{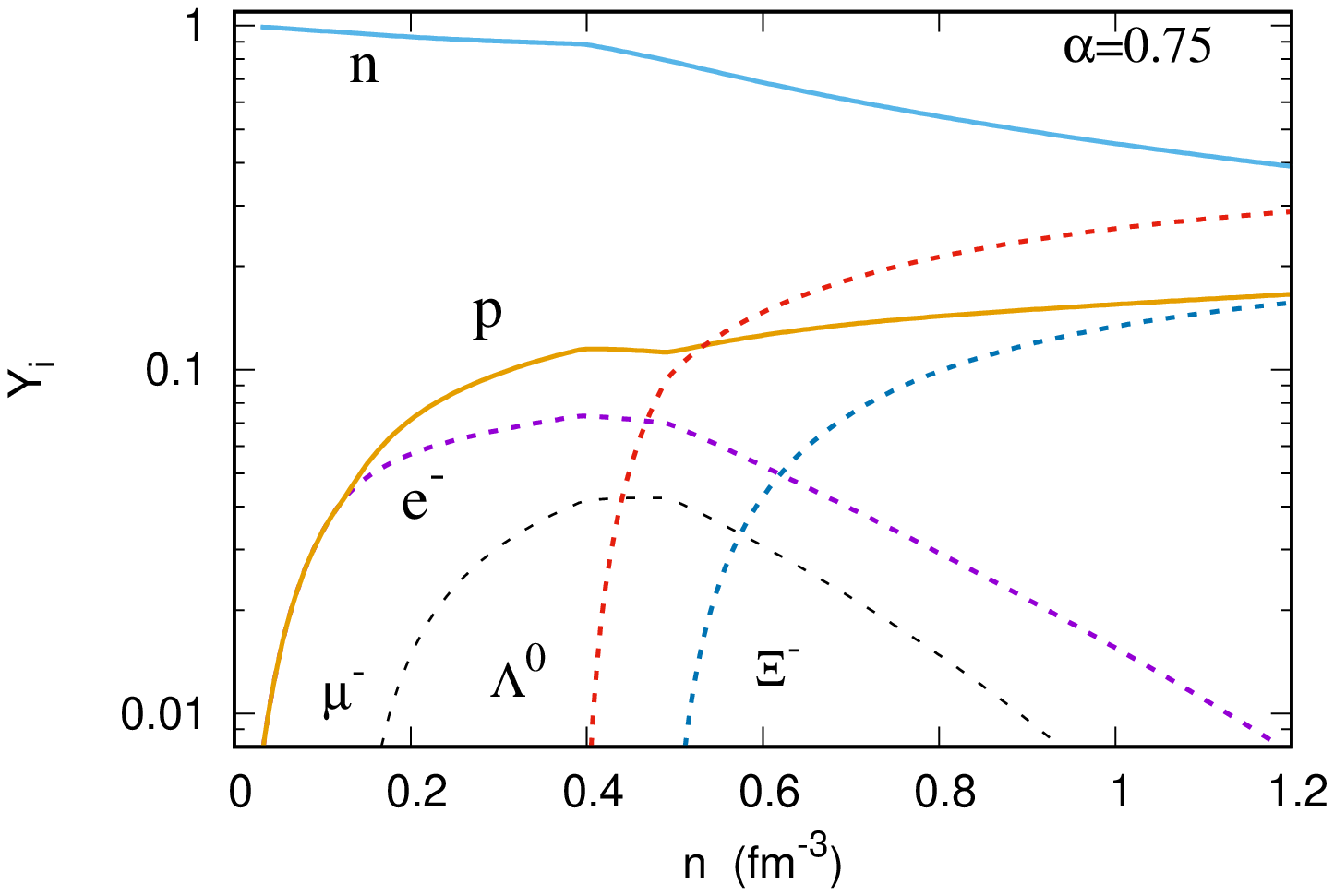}

\includegraphics[width=.3\textwidth, angle=270]{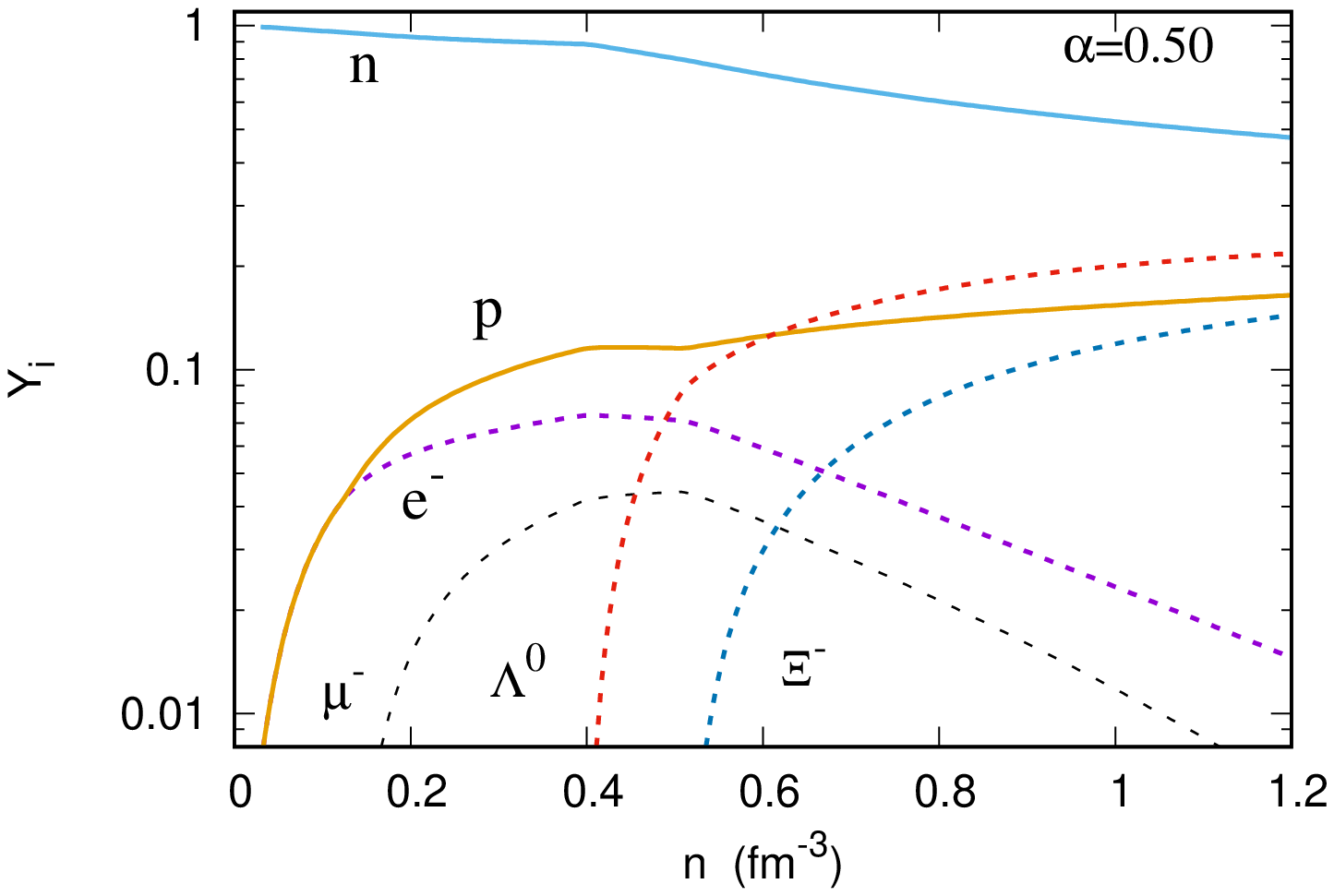} 
\includegraphics[width=.3\textwidth, angle=270]{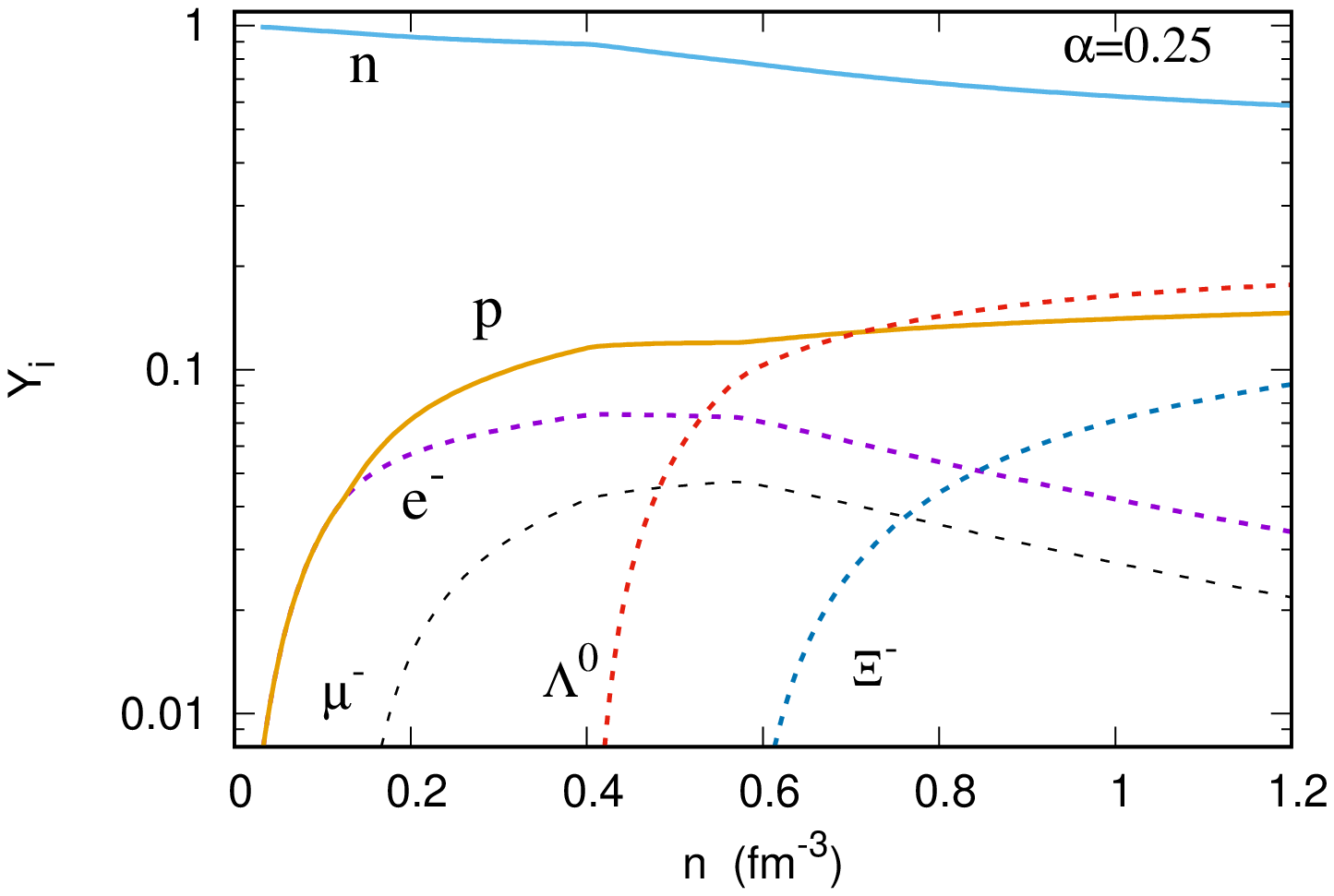}
\caption{(Color online) Particle population for different values of $\alpha_v$ for hyperonic neutron star matter. } \label{FL1}
\end{figure*}

We plot in Fig.~\ref{FL1} the particle population for different values of $\alpha_v$. As can be seen, the $\Sigma$ triplet is not present in any parameterization, due to the highly repulsive potential - $U_\Sigma = +30$ MeV - together with the strong $\omega$ repulsion, especially for low values of $\alpha_v$.
For the SU(6) parameterization ($\alpha_v=1.00$), the lepton fraction goes to zero at high densities, and the charge neutrality is obtained by an equal fraction of protons and $\Xi^-$ hyperons.
As we decrease $\alpha_v$, the lepton fraction at high densities increases. In all parameterizations we also see that the $\Lambda^0$ is the first hyperon to appaer, always around $n=0.40$ fm$^{-3}$.

\begin{figure}[!t] 
\begin{centering}
 \includegraphics[angle=270,
width=0.46\textwidth]{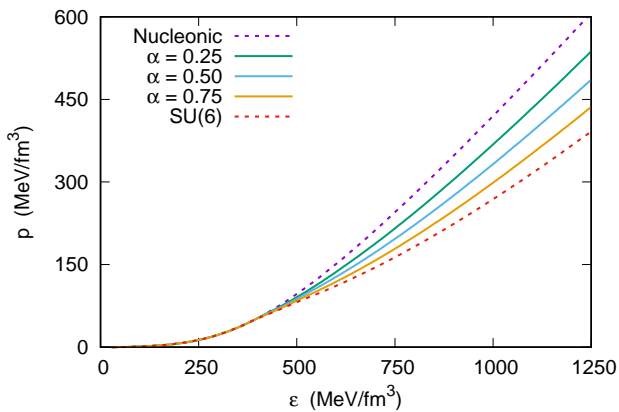}
\caption{(Color online)  Equation of state for different values of $\alpha_v$ for hyperonic nuclear matter.} \label{FL2}
\end{centering}
\end{figure}

We plot in Fig.~\ref{FL2} the EOS for hyperonic nuclear matter for different values of $\alpha_v$.
We can see, that there is a clear relation between $\alpha_v$ and the stiffness of the EOS. The lower the value of $\alpha_v$, the smaller the hyperon fraction present in the matter composition and so, as stated by the hyperon puzzle reasoning, the stiffer the EOS. Such result was already pointed in Refs.~\cite{Lopes2013,lopesnpa,Weiss2}.

\begin{figure}[!t] 
\begin{centering}
 \includegraphics[angle=270,
width=0.46\textwidth]{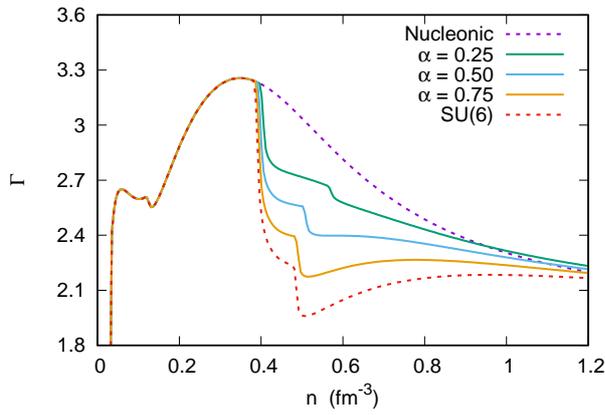}
\caption{(Color online) The adiabatic index  for different values of $\alpha_v$ for hyperonic nuclear matter.} \label{FL3}
\end{centering}
\end{figure}

The adiabatic index, defined as:
\begin{equation}
\Gamma = \frac{(p +\epsilon)}{p}\left( \frac{\partial p}{\partial \epsilon} \right), \label{eadia}
\end{equation}
is a sensitive indicator of phase changes in stellar matter and the stability with respect to vibrations and pulsations of the star~\cite{Haensel2002,Haensel2008}. 
For multicomponent matter, the adiabatic index exhibit jumps at densities coincident with density thresholds of individual components, signaling phase transitions and/or changes in the matter constitution~\cite{Stone2021}. 
The adiabatic index $\Gamma$ presents information not only on the EOS ($p$ and $\epsilon$), but also on the speed of sound ($v_s^2 =  \partial p/\partial \epsilon$). It is shown in Fig.~\ref{FL3}.
Looking at Fig.~\ref{FL3}, one can notice that the $\Lambda^0$ threshold around 0.40 $fm^-3$ is strongly evidenced, causing a
huge drop in the value of $\Gamma$. A smaller but distinguishable peak also points to the onset of the $\Xi^-$ hyperon. Also, the higher the $\alpha_v$, the deeper the drop of the adiabatic index due to the $\Lambda^0$ threshold.

\begin{figure}[!t] 
\begin{centering}
 \includegraphics[angle=270,
width=0.46\textwidth]{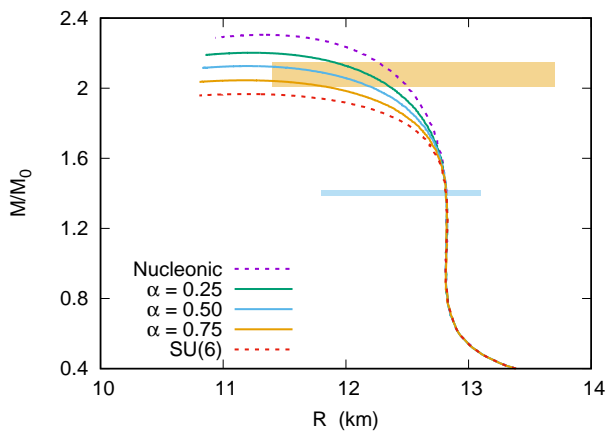}
\caption{(Color online) Mass-radius relation for hyperonic neutron stars for different values of $\alpha_v$. The hatched areas correspond to the uncertainty on the radius of the canonical star, and the mass-radius uncertainty of the PSR J0740+6620 pulsar.  } \label{FL4}
\end{centering}
\end{figure}

Now  we turn our attention to the neutron star properties that result from each EOS. Any valid EOS must be able to reproduce some astrophysical constraints. Maybe the more important is that it must be able to describe the mass and radius of the PSR J0740+6620 pulsar, whose  values are $M=2.08\pm 0.07$ M$_\odot$ and $R=12.39_{-0.98}^{+1.30}$ km \cite{NICER3}. Another important constraint is the radius of the canonical $M=1.4$ M$_\odot$ star. We use here a very strong constraint presented in  Ref.~\cite{Miller}, which suggest $R_{1.4} = 12.45 \pm 0.65$ km, that limits the radius of the canonical star within an uncertainty of only 5$\%$. We also discuss the
possibility of the existence of a very massive neutron star, dubbed as \textit{black widow}, with a mass of $M=2.35 \pm 0.17$ M$_\odot$, as suggested in Ref.~\cite{Romani2022}.
Such heavy star, if confirmed, would be by far the most massive  neutron star ever measured.

The mass-radius relations are obtained by using the EOS as input to solve the TOV equations~\cite{TOV}. Complementary,  we use the BPS EOS~\cite{BPS} for the outer crust  and the BBP EOS~\cite{BBP} for the inner crust, as discussed in Ref.~\cite{Haenselbook}. The results are presented in Fig.~\ref{FL4}.
As can be seen, with the exception of the SU(6) parameterization, all models are able to describe the $M= 2.08 \pm 0.07$  M$_\odot$ PSR J0740+6620 pulsar. Moreover, as the hyperon onset is around $n=0.4$ fm$^{-3}$, within our model only stars with masses higher than $M=1.43$ M$_\odot$  present hyperons in their core. Given that, all canonical stars has the same internal nucleon-only composition and, therefore the same radius of $R=12.82$ km. This result is in agreement with the discussion presented in Ref.~\cite{Miller}. Finally, for $\alpha_v = 0.25$, we found a maximum mass of $M=2.20$ M$_\odot$, what means that this model allows the description of the PSR J0740+6620 as containing hyperons in its the core, and even of the  black widow pulsar in the lower limit of the error bar.

\begin{figure}[!t] 
\begin{centering}
 \includegraphics[angle=270,
width=0.46\textwidth]{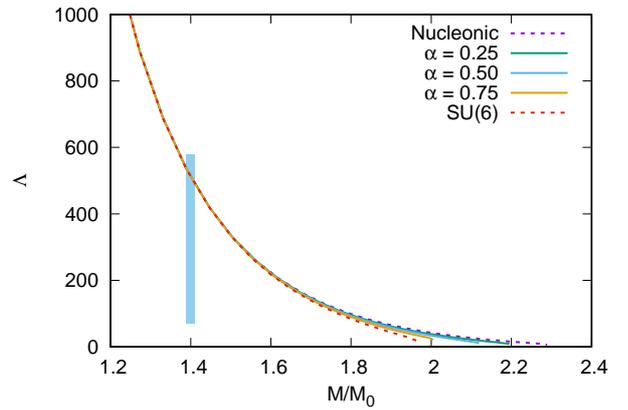}
\caption{(Color online) Dimensionless tidal deformability parameter $\Lambda$ for  hyperonic neutron stars for different values of $\alpha_v$. The hatched area corresponds to the uncertainty on the $\Lambda_{1.4}$ value obtained from the GW170817 event~\cite{tidal1}.  } \label{FL5}
\end{centering}
\end{figure}

Another important astrophysical constraint comes from the GW170817 event, detected by the LIGO/VIRGO gravitational wave telescopes: the dimensionless tidal deformation parameter $\Lambda$.
The tidal deformability of a compact object is a single parameter that quantiﬁes how easily the object is deformed when subjected to an external gravitational ﬁeld. Larger tidal deformability indicates that the object is easily deformable. On the opposite side, a compact object with a smaller tidal deformability parameter is smaller, more compact, and it is more difﬁcult to deform. Its is deﬁned as:

\begin{figure*}[!t]
\centering %
\includegraphics[width=.3\textwidth, angle=270]{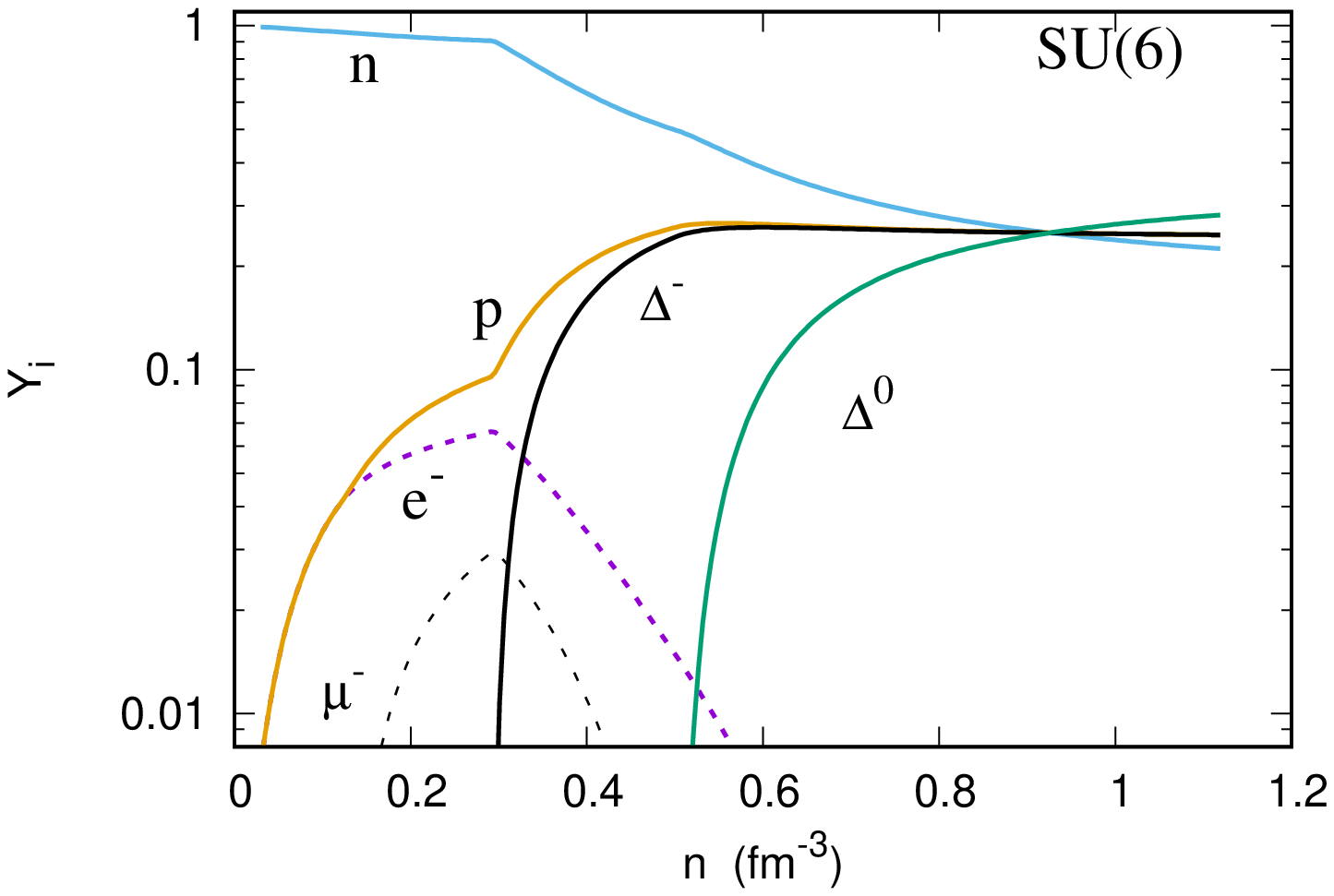}
\includegraphics[width=.3\textwidth, angle=270]{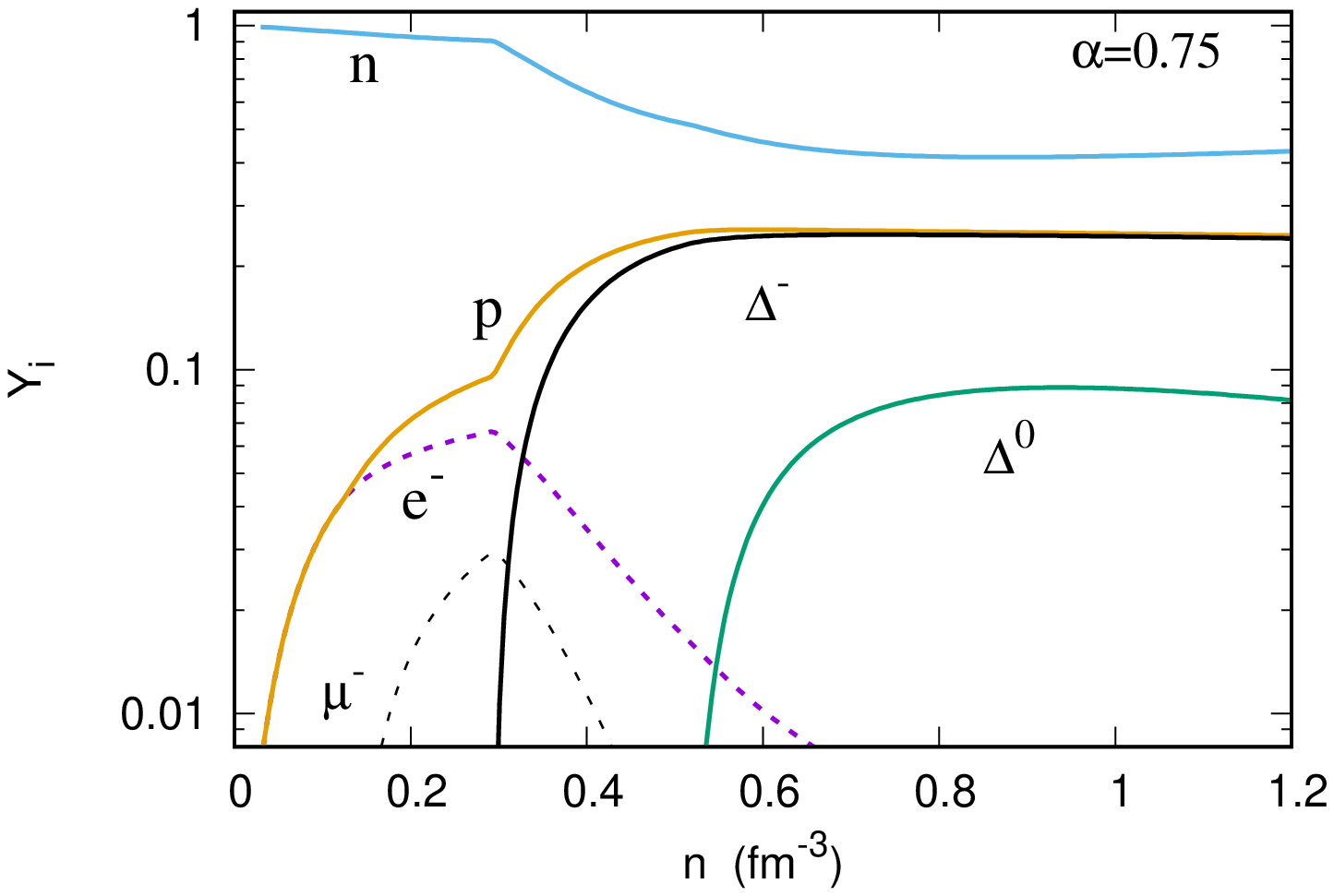}

\includegraphics[width=.3\textwidth, angle=270]{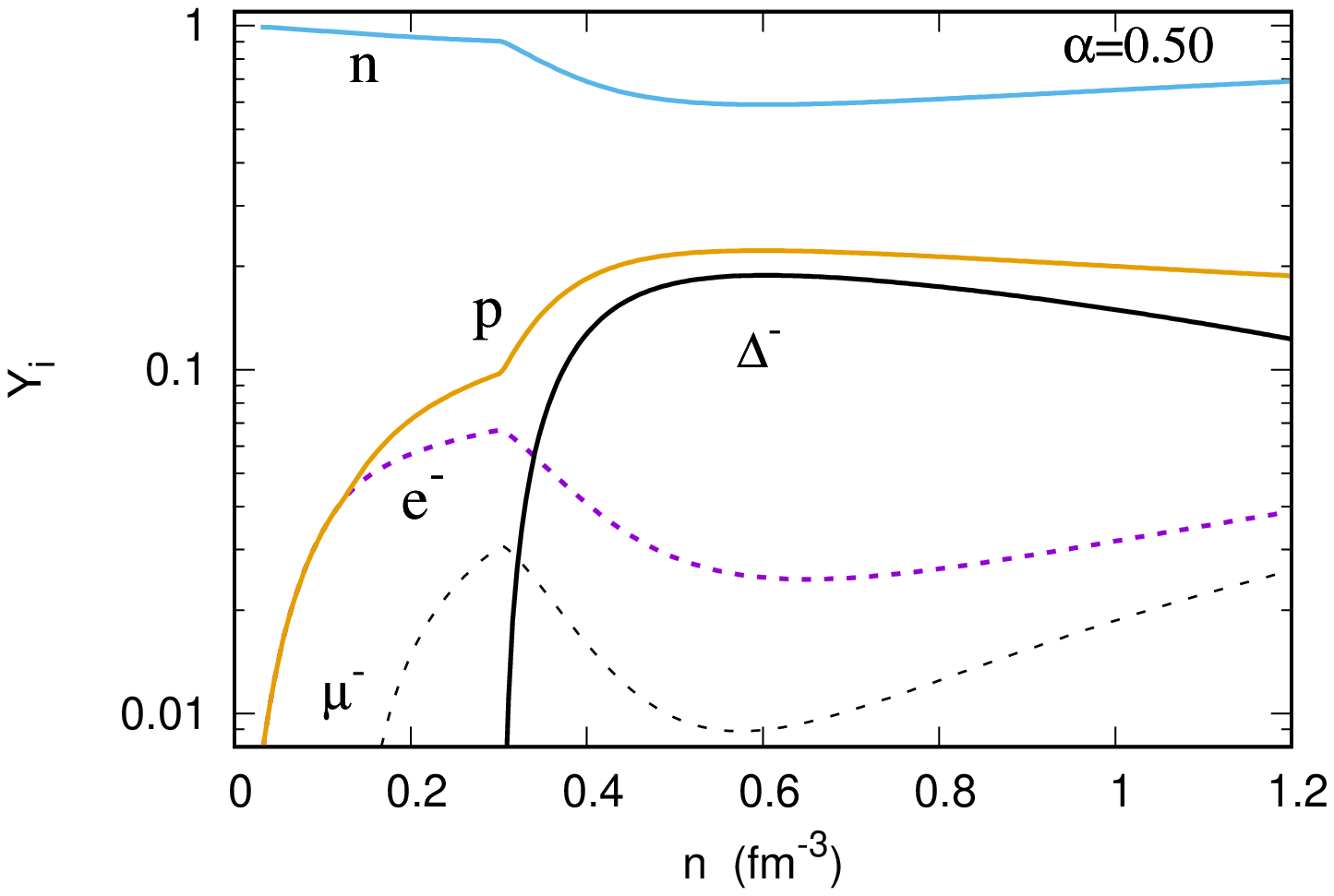}
\includegraphics[width=.3\textwidth, angle=270]{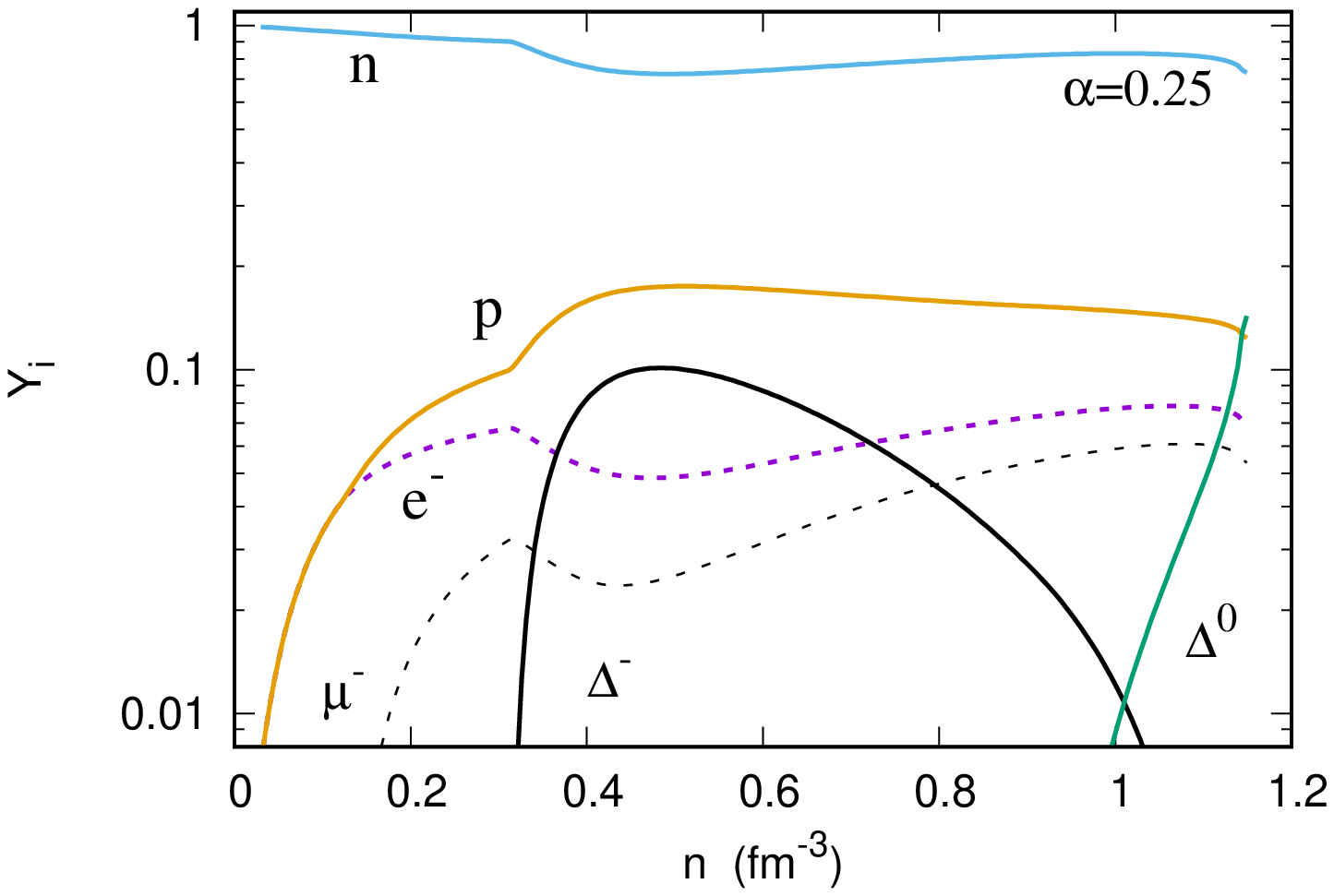}
\caption{(Color online) Particle population for different values of $\alpha_v$ for $\Delta$-admixed neutron star matter. } \label{FL6}
\end{figure*}

\begin{equation}
\Lambda = \frac{2k_2}{3C^5}, \label{tidal}
\end{equation}
where $C = GM/R$ is the compactness of the star. The parameter $k_2$ is called the Love number and is related to the metric perturbation. We refer the interested reader to Refs.~\cite{tidal1,tidal2,tidal3,tidal4} to a complete discussion about the Love number and its calculation procedure. Here, we adopt the constraint on the dimensionless tidal parameter for the canonical mass star presented in Ref.~\cite{tidal1}, ${70\leq\Lambda_{1.4}\leq580}$. As there are no hyperons in the 1.4 M$_\odot$ star, all  values of $\alpha_v$ produce the same dimensionless tidal parameter $\Lambda_{1.4}=516$, which is in agreement with this constraint. The results for the  tidal deformability are presented in Fig.~\ref{FL5}. Finally, the hyperonic neutron star main properties are presented in Tab.~\ref{T3}.

\begin{table}[!t]
\begin{center}
\begin{tabular}{c|cccc|c}
&\multicolumn{4}{c|}{$\alpha_v$}&{ }\\
 & SU(6) & 0.75 &  0.50 &  0.25  & Nucl. \\
\hline
 M$_{\rm max}$/M$_\odot$      & 1.97  & 2.05   & 2.13 & 2.20 & 2.30   \\
 $n_c$ (fm$^{-3}$)        & 1.04 & 1.02  & 1.00 & 0.98 & 0.94  \\
R  (km)       & 11.18 & 11.17  & 11.19 & 11.23 & 11.34  \\
R$_{1.4}$  (km)          & 12.82 & 12.82  & 12.82 & 12.82 & 12.82  \\
$\Lambda_{1.4}$         & 516 & 516  & 516 & 516 & 516 \\
\hline
\end{tabular}
 \caption{Hyperonic neutron star main properties, for the maximum mass and canonical stars. The properties of the nucleonic star is included for comparison. 
 }\label{T3}
 \end{center}
 \end{table}

\section{$\Delta$-admixed neutron stars}

Now we study the effects of the presence of $\Delta$ baryons on the neutron star properties.
We must emphasize that here we consider only the 
$\Delta^-$ and $\Delta^0$ baryons, because the inclusion of the $\Delta^+$ and $\Delta^{++}$ baryons causes the nucleon  effective mass to drop to zero for very low densities, preventing the neutron stars to reach densities high enough to describe in the maximum mass star. This behaviour was discussed thoughtfully in Ref. \cite{nossoConstanca}, where it is understood that the increase of the exotic particle abundance adds to the negatively contributing term of the effective nucleon mass, through the scalar density dependence of the $\sigma$ field. In that study, the authors could tune the multiplicity of baryons in the matter by varying the coupling constants, what would go against the scope of the present work. Yet, as the couplings are now bound by the symmetry relations, determining the onset of the particles, the abundance of positively-charged baryons in the lower densities makes it very hard to obey the charge neutrality condition.

We show  the particle population for $\Delta$-admixed neutron star matter in Fig.~\ref{FL6}. As it can be seen, the $\Delta^-$ is always the first one to appear, at a density about $n=0.3$ fm$^{-3}$.
This can be explained due to the high attractive potential of the $\Delta$'s ($U_\Delta = -98$ MeV) as well as its negative electric charge, which reduces its chemical potential. Another fact is that, since the members of the spin 3/2 decuplet carry a $\gamma=4$ degeneracy factor, their number densities are twice bigger than it would be for a spin 1/2 baryon. 
This makes the increase of the $\Delta^-$ population very fast. 
For SU(6) and for $\alpha_v = 0.75$, the density of the $\Delta^-$ quickly reaches the density of the protons, contributing to the deleptonization of the nuclear matter because the charge neutrality is now reached by an equal amount of protons and $\Delta^-$. For the SU(6) case, the neutron population is outweighed by all the other baryonic species, with $\Delta^0$ becoming the more abundant component in the high density region. For SU(6) and $\alpha_v = 0.75$, the $\Delta^0$ appears before $n=0.6$ fm$^{-3}$, being absent for $\alpha_v = 0.50$, but reappearing for $\alpha_v = 0.25$ at very high density.
Seeing beyond the model-dependency specifics, we can compare these results with the ones presented in Ref. \cite{nossoConstanca}. There, it is stated that the larger the difference between the vectorial (repulsive) and the scalar (attractive) couplings, the less favored the $\Delta$ baryons
are to appear. The overall behavior of the particle population is consistent between the two studies, with the main difference being that, in the SU(6) case here, the $\Delta^0$ population dominates in the higher density, while in \cite{nossoConstanca} the  $\Delta^-$ is always more abundant. It can be explained by the absence of the positively-charged resonances here, whose onset would increase the $\Delta^-$ population.
Also, we see that for $\alpha_v= 0.50$  and 0.25, the $\Delta^-$ fraction begins to decrease at high densities, which causes an increase of the lepton fraction. Such behaviour has never been  observed in hyperonic nuclear matter.
{This behavior is due to the fact that resonances are subject to a coupling with the mesons stronger than the other baryons. It favors their onset at low densities, where the attractive part of the potential is more relevant, but makes them less favored once  the repulsive $\omega$ field dominance takes place at large densities that occurs because the $\sigma$ field saturates \cite{nossoConstanca}.}

\begin{figure}[!t] 
\begin{centering}
 \includegraphics[angle=270,
width=0.46\textwidth]{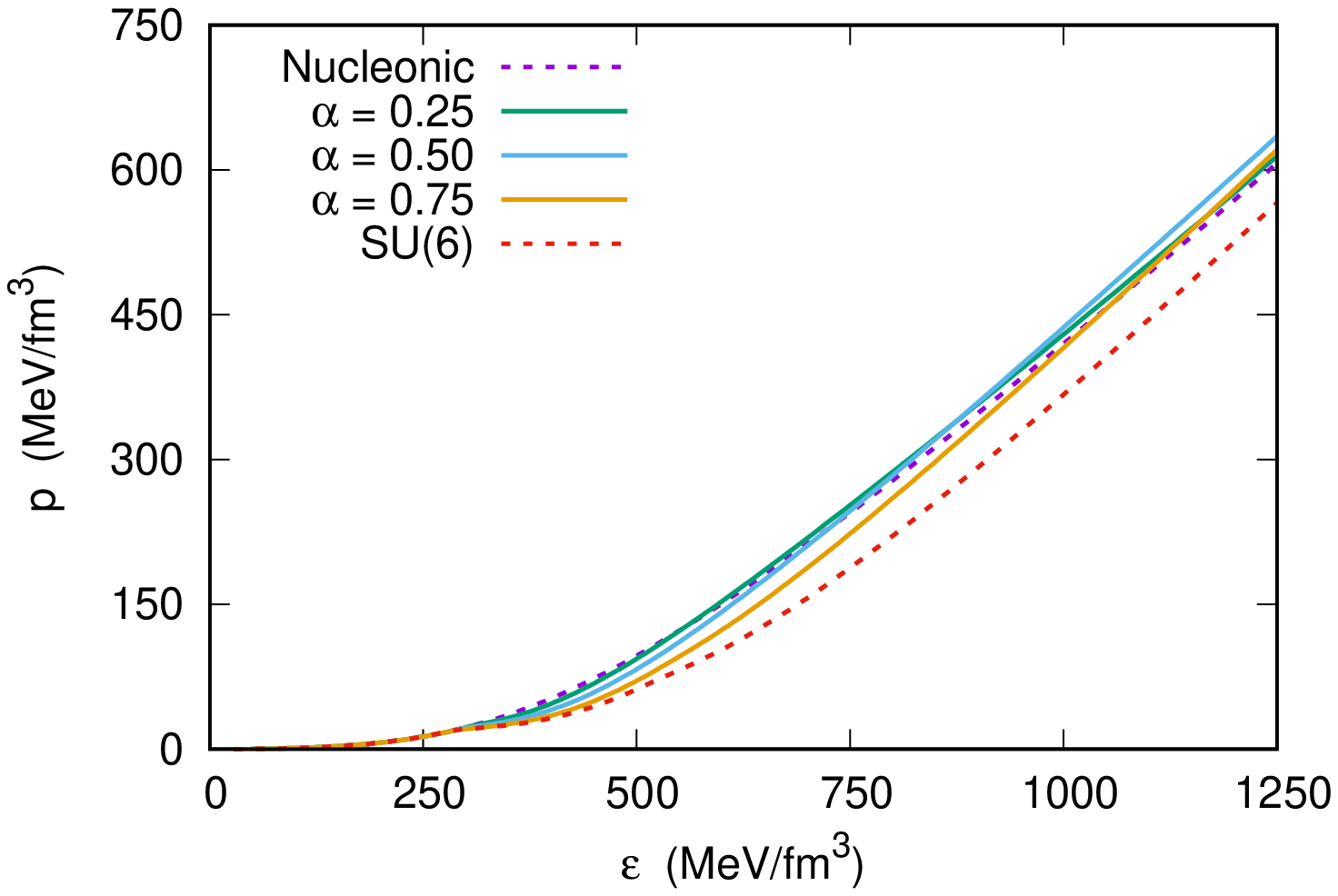} 
\includegraphics[angle=270,
width=0.46\textwidth]{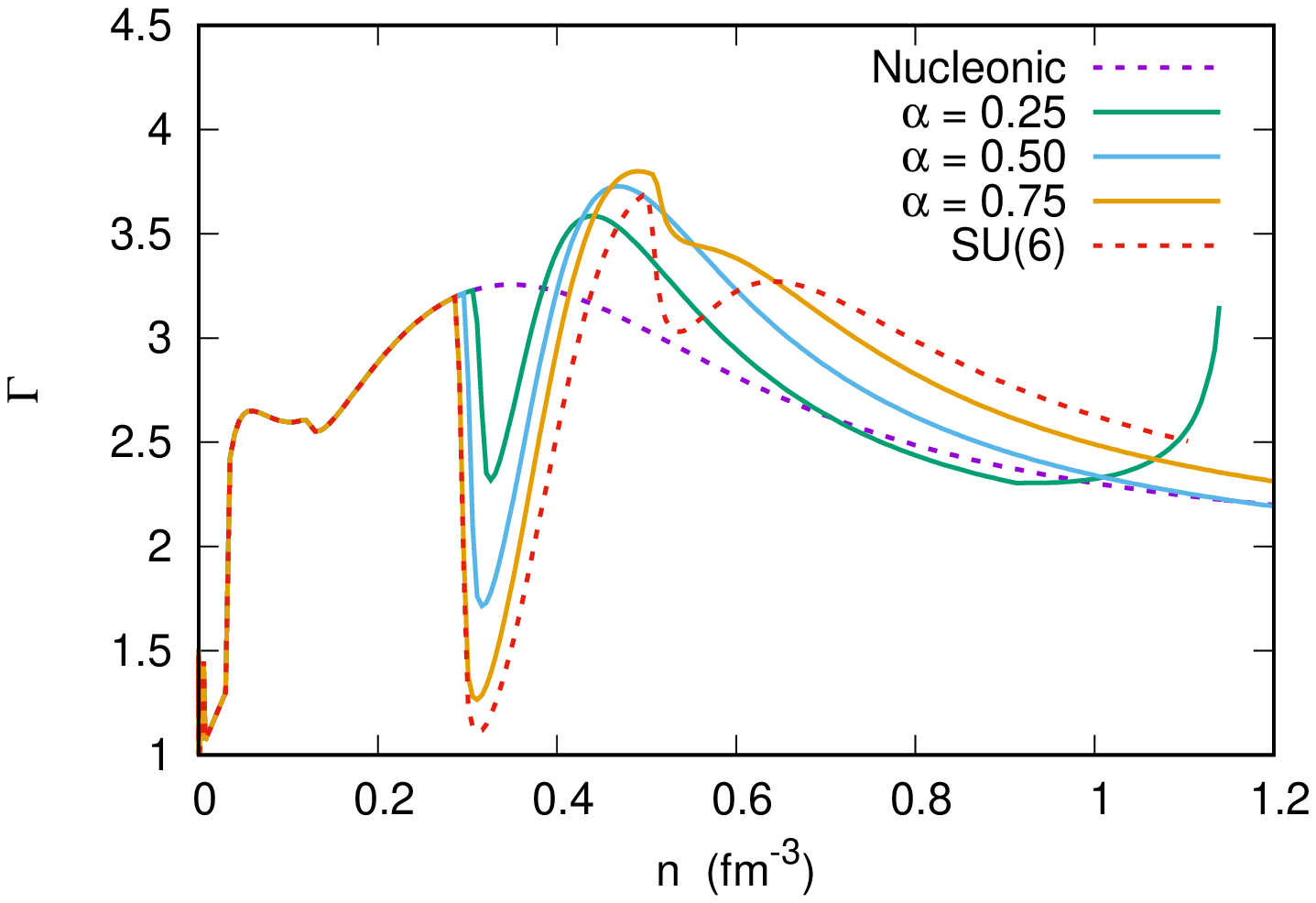}
\caption{(Color online) Equation of state (top) and adiabatic index $\Gamma$ (bottom) for $\Delta$-admixed neutron star matter for different values of $\alpha_v$.  } \label{FL7}
\end{centering}
\end{figure}

We now study how the $\Delta$ threshold affects the nuclear EOS and the adiabatic index $\Gamma$, both results presented in Fig.~\ref{FL7}. For the SU(6) and $\alpha_v = 0.75$, the EOS is clearly softer than the pure nucleonic one. But for $\alpha_v = 0.50$ and 0.25, this is not so clear. In fact for these values the EOS becomes stiffer than the pure nucleonic one at high densities. Again, let's point out that such behaviour is never present in the case of hyperonic nuclear matter. All the complexity of the $\Delta$-admixed nuclear matter is more strongly reflected in the adiabatic index. There is a huge drop of $\Gamma$ around $n=0.3$ fm$^{-3}$, due to the onset of the $\Delta^-$, 
which is  similar to the drop caused by the $\Lambda^0$ threshold but  deeper. What is not present in the case of the hyperonic nuclear matter is the huge and fast increase of the $\Gamma$ around $n=0.4$ fm$^{-3}$. Indeed, the adiabatic index becomes higher for $\Delta$-admixed nuclear matter than for pure nucleonic matter. For $\alpha_v=$  0.25, we also see a new increase of the $\Gamma$ due to the onset of the $\Delta^0$, just before the numerical code stops converging.

\begin{figure}[!h] 
\begin{centering}
 \includegraphics[angle=270,
width=0.46\textwidth]{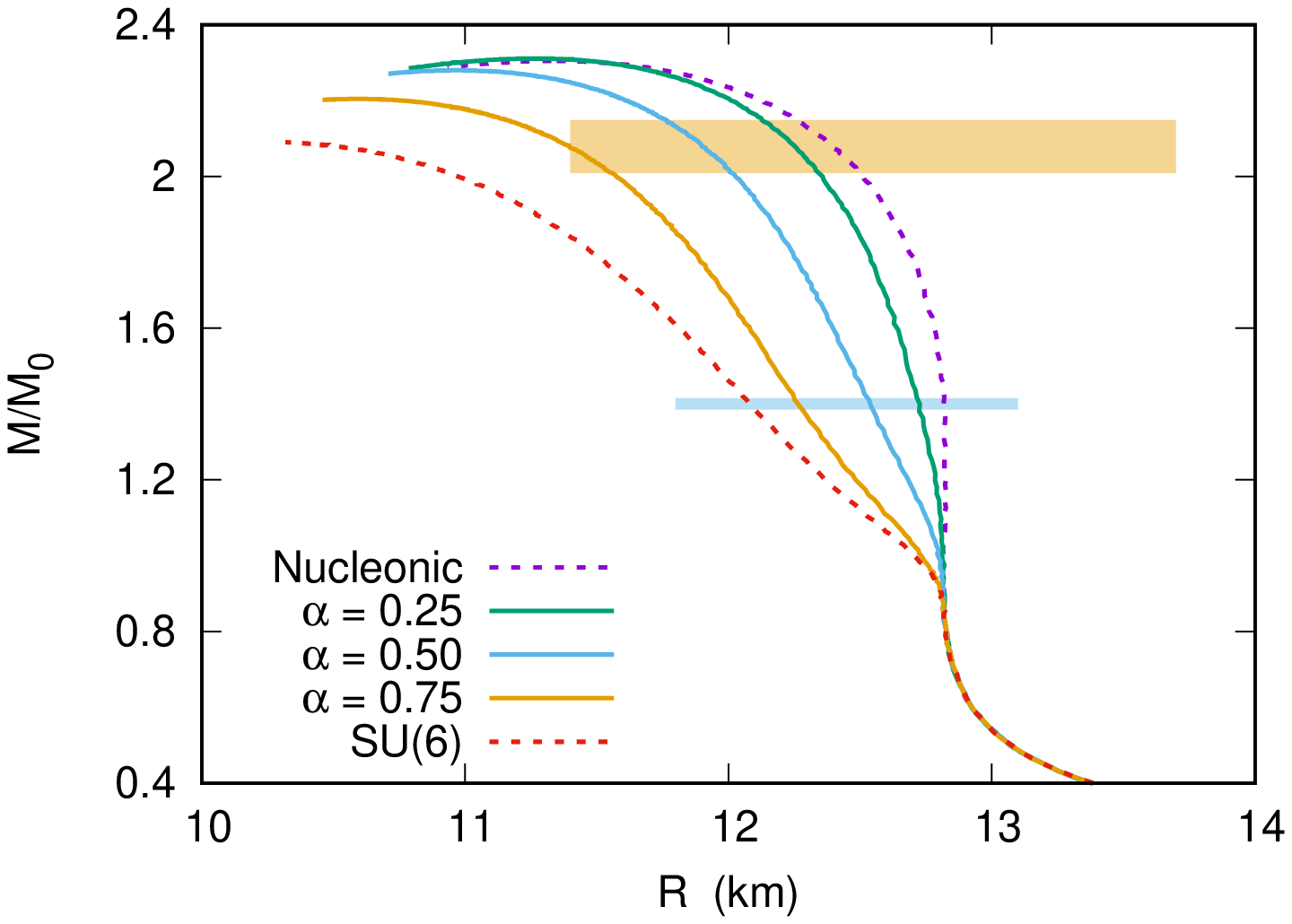} \\
\includegraphics[angle=270,
width=0.46\textwidth]{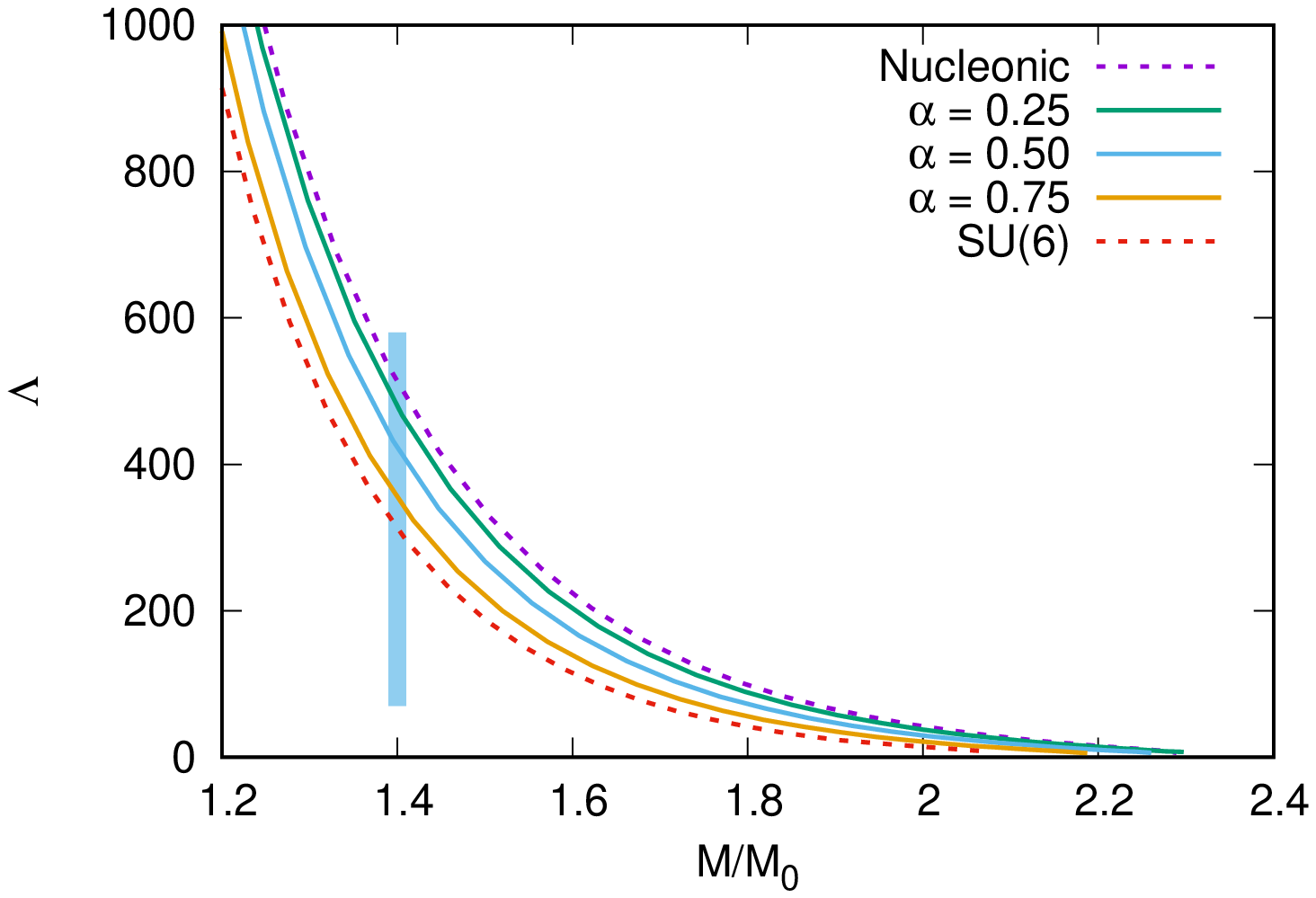}
\caption{(Color online) TOV solutions (top) and dimensionless tidal parameter $\Lambda$ (bottom) for $\Delta$-admixed neutron stars for different values of $\alpha_v$.  } \label{FL8}
\end{centering}
\end{figure}

Solving the TOV equations we can see how the onset of the $\Delta'$s affects the neutron star macroscopic properties and the dimensionless tidal parameter, as presented in Fig.~\ref{FL8}. As the $\Delta^-$ threshold happens at $n=0.3 $ fm$^{-3}$ this implies that stars with masses above $M=0.92 $ M$_\odot$ contain $\Delta$'s in their core. In practice, this means that all known neutron stars have deltas in their cores if we assume a $\Delta$-admixed neutron star matter composition. As a direct consequence, the radius of the
canonical star depends on the $\Delta$-meson coupling constants. We see that the higher the value of $\alpha_v$, the lower the maximum mass and the lower the radius of the canonical star. The SU(6) parameterization is the only one which is not able to describe  the PSR J0740+6620 pulsar~\cite{NICER3}. Although the maximum mass is $M=2.09 $ M$_\odot$, this curve misses  the observational data window by presenting radii below $R=11.40$ km. For all other values of $\alpha_v$ we are able to explain the PSR J0740+6620 pulsar, and for $\alpha_v = 0.50$ and 0.25, even a mass above 2.18 $M_\odot$  is reached, explaining the PSR J0952-0607 black widow pulsar discussed in Ref.~\cite{Romani2022}. A very interesting feature here is the fact that for $\alpha_v = 0.25$, the
maximum mass is $M=2.31$ M$_\odot$, even larger than the maximum mass of a pure nucleonic neutron star of $M=2.30$ M$_\odot$.
Again, this behavior is due to the fact that the resonances are subject to a stronger coupling with the mesons than the nucleons. 
Larger scalar coupling favors larger $\Delta$ fractions and, as the deltas also couple more strongly to the vector fields, the $\omega$-dominance at large densities results in stiffer EOS, and, therefore, larger masses, as discussed in Ref. \cite{nossoConstanca}.

As far as the radius of the canonical star is concerned, we see that all parameterizations are in agreement with the discussion presented in Ref.~\cite{Miller}. Indeed, for SU(6), the canonical radius can be as low as $R=12.08$ km. The presence of $\Delta$'s in the neutron star core can naturally explain why some observational astrophysical results point to a very low radius for the canonical star~\cite{Capano}, while at the same time some terrestrial nuclear experiments point to a high slope of the nuclear matter symmetry energy~\cite{PREX2}, that would suggest larger radii.  The increase of the neutron star compactness due to the onset of a new degree of freedom was already discussed in Ref.~\cite{Lopes2018}.   About the dimensionless tidal parameter, we see that the increase of the compactness of the star due to the onset of $\Delta^-$ also reduces the $\Lambda_{1.4}$. For SU(6), this value is as low as $\Lambda_{1.4}=318$. The $\Delta$-admixed neutron stars main properties are summarized in Tab.~\ref{T4}.

\begin{table}[!t]
\begin{center}
\begin{tabular}{c|cccc|c}
&\multicolumn{4}{c|}{$\alpha_v$}&{ }\\
 & SU(6) & 0.75 &  0.50 &  0.25  & Nucl. \\
\hline
 M$_{\rm max}$/M$_\odot$      & 2.09  & 2.20   & 2.28 & 2.31 & 2.30   \\
 $n_c$ (fm$^{-3}$)        & 1.11 & 1.06  & 0.99 & 0.95 & 0.94  \\
 R  (km)       & 10.31 & 10.59  & 10.97 & 11.26 & 11.34  \\
 R$_{1.4}$  (km)          & 12.08 & 12.25  & 12.53 & 12.73 & 12.82  \\
$\Lambda_{1.4}$         & 318 & 360  & 428 & 489 & 516 \\
\hline
\end{tabular}
 \caption{$\Delta$-admixed neutron star main properties, for the maximum mass and canonical stars. The properties of the nucleonic star is included for comparison..}\label{T4}
 \end{center}
 \end{table}

\section{$\Delta$-admixed hyperonic neutron stars}

\begin{figure*}[!t]
\centering
\includegraphics[width=.3\textwidth, angle=270]{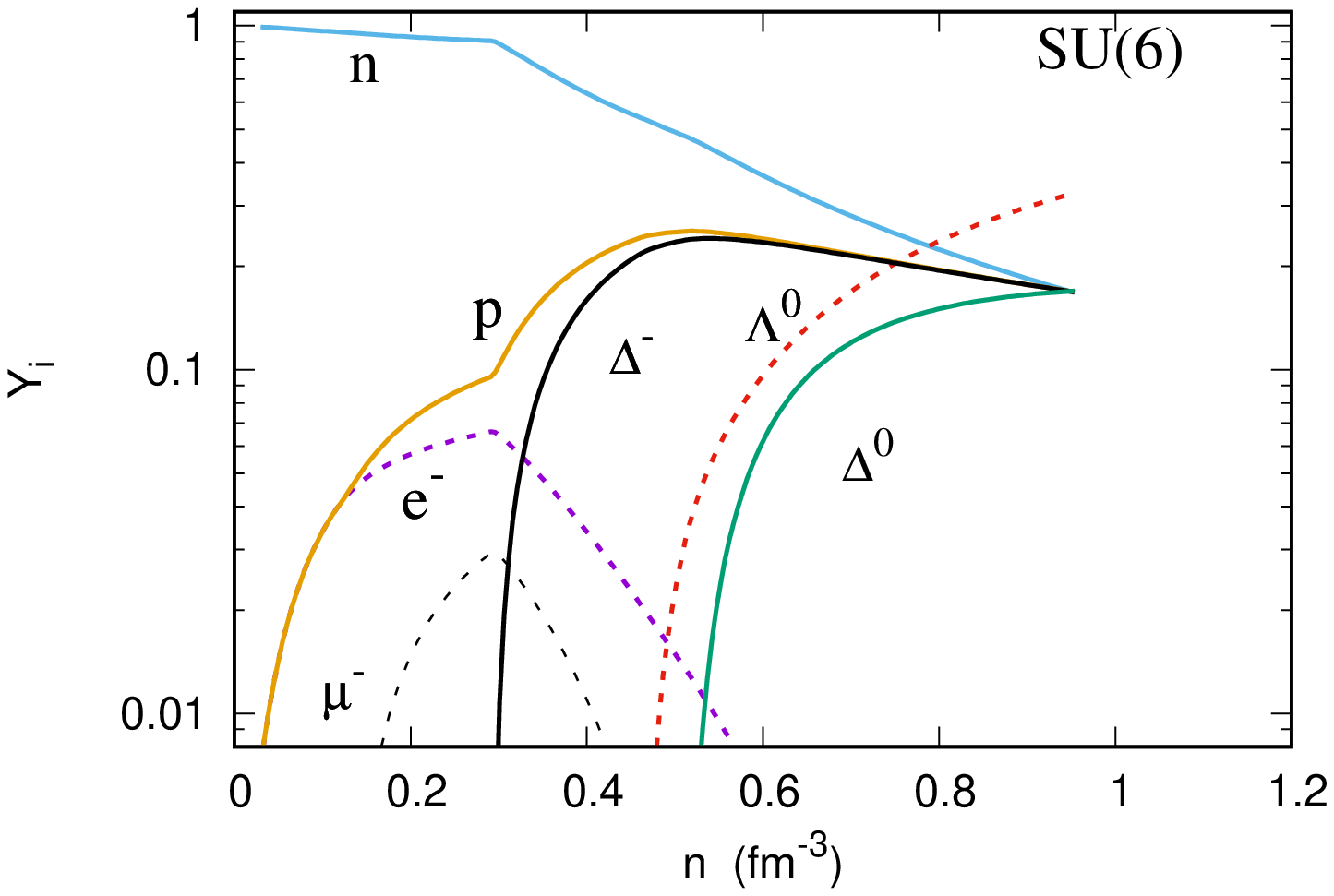} 
\includegraphics[width=.3\textwidth, angle=270]{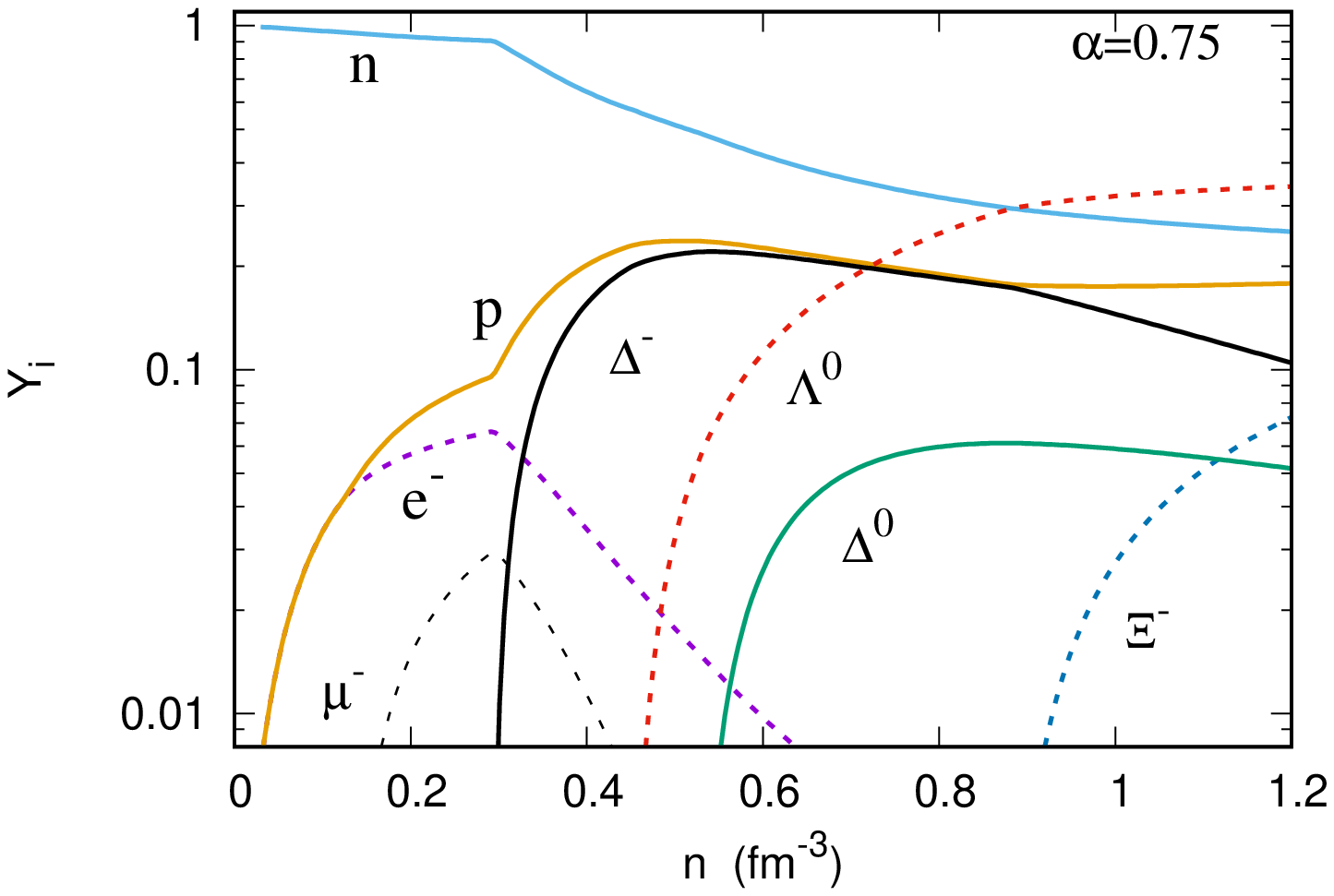}

\includegraphics[width=.3\textwidth, angle=270]{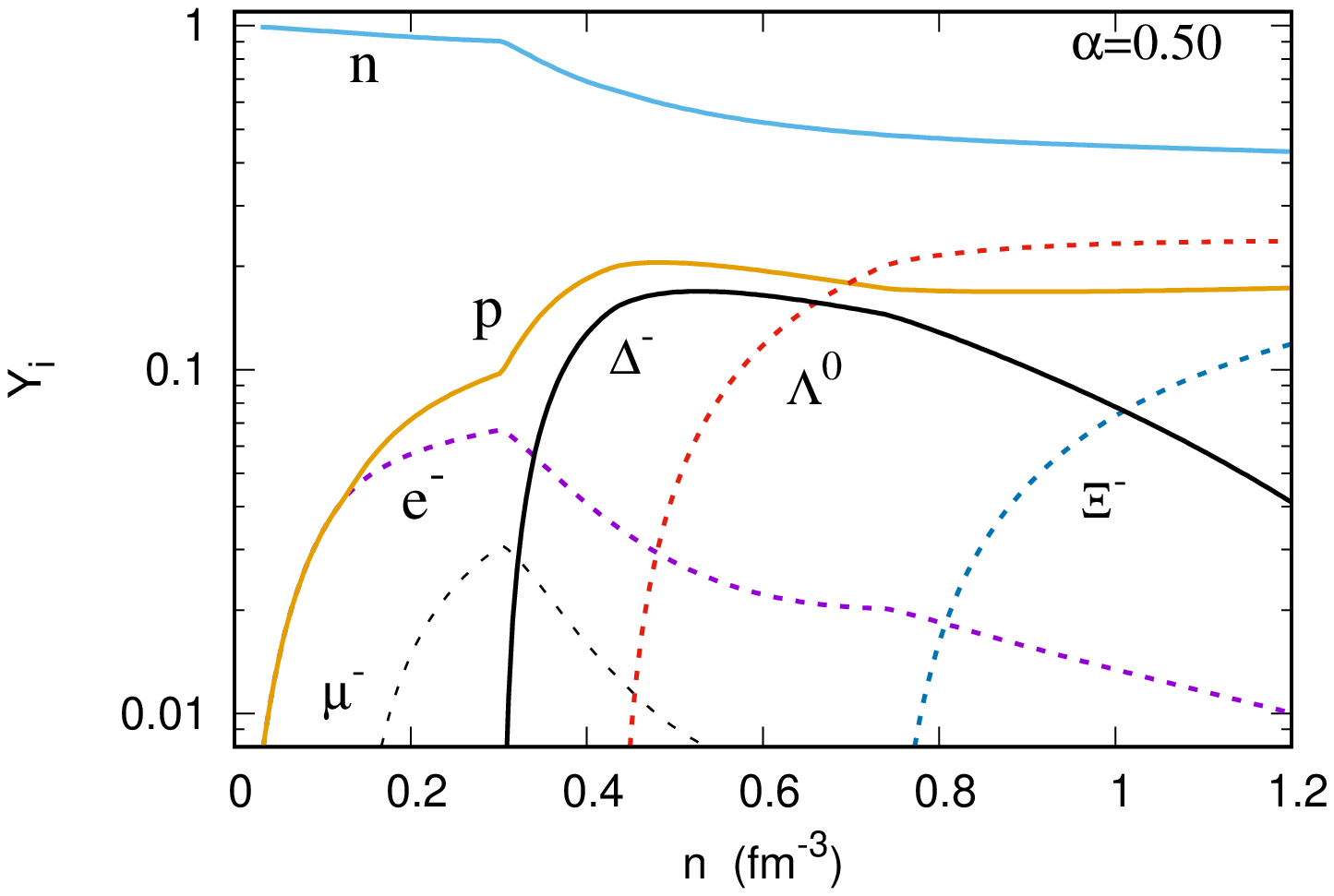} 
\includegraphics[width=.3\textwidth, angle=270]{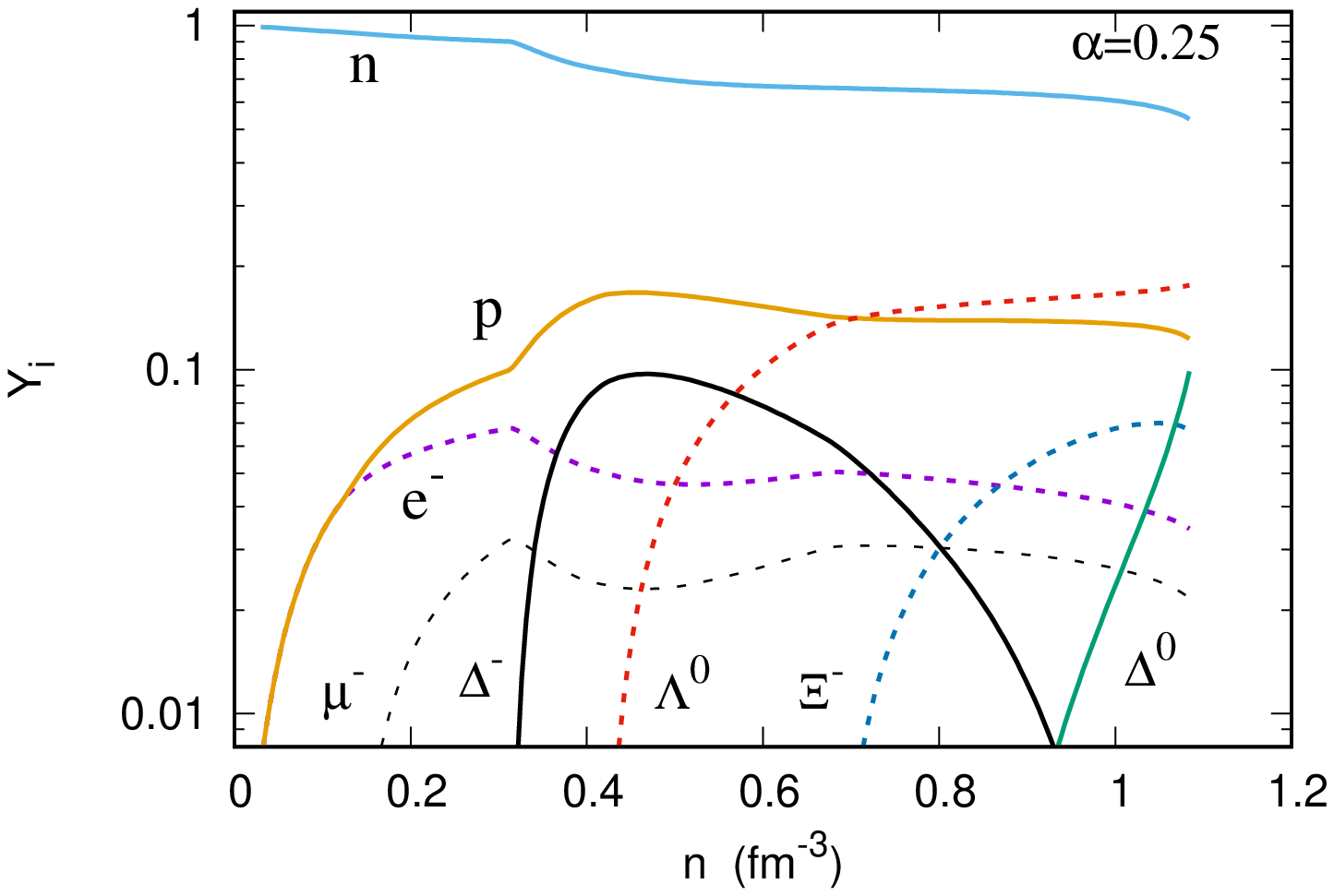}
\caption{(Color online) Particle population for different values of $\alpha_v$  for $\Delta$-admixed hyperonic matter. } \label{FL9}
\end{figure*}

In this last section we study how the presence of both $\Delta$ barryons and hyperons affect the nuclear EOS and the corresponding neutron star macroscopic properties. In this case there is expected a very complex competition between the $\Delta$'s and the hyperons populations. The particle populations are presented in Fig.~\ref{FL9}.
We can see that the onset of the $\Delta^-$ around $n=0.3$ fm$^-$ is kept, as in the case of when only $\Delta$'s are admixed to the neutron star matter. On the other hand, the $\Lambda^0$ has its threshold pushed away towards higher densities when compared with pure hyperonic nuclear matter, due to the higher attractive potential of the resonances. For the SU(6), the $\Xi^-$ is not present, once the nucleon mass goes to zero before its onset and the numerical code stops converging. As in the case of pure $\Delta$-admixed matter, for $\alpha_v = 0.50$, the $\Delta^0$ is not present.
For $\alpha_v = 0.50$ and 0.25, the releptonization of the star matter due to the decrease of the $\Delta^-$ seen in the pure $\Delta$-admixed matter is now weakened, once there is the onset of the $\Xi^-$ in these scenarios.
It is more difficult to draw parallels between our results here and the ones in Ref. \cite{nossoConstanca} because the authors there considered a fixed SU(6) hyperon coupling scheme and varied the delta couplings freely. However, apart from model dependencies and the suppression of the positively-charged deltas, some behaviors repeat themselves in both studies. The $\Delta^-$ is the first exotic particle to appear. The reasoning to that is that it can replace a neutron-electron pair at the top of their Fermi seas, being favored
over the lighter hyperons because their potential is more attractive.  The $\Lambda^0$, being electrically neutral and the lighter exotic baryon, is the next to appear. It also becomes the more abundant particle in intermediate densities, after the $\omega$-dominance gets estabilished, because its smaller repulsive coupling.
The competition between the $\Delta^-$ and the $\Xi^-$  can also be understood by the former being subject to a less repulsive coupling than the first.
Also, we argue that there is a strong tendency of deleptonization in the high density region of $\Delta$-admixed hyperonic neutron stars.

\begin{figure}[!t] 
\begin{centering}
 \includegraphics[angle=270,
width=0.46\textwidth]{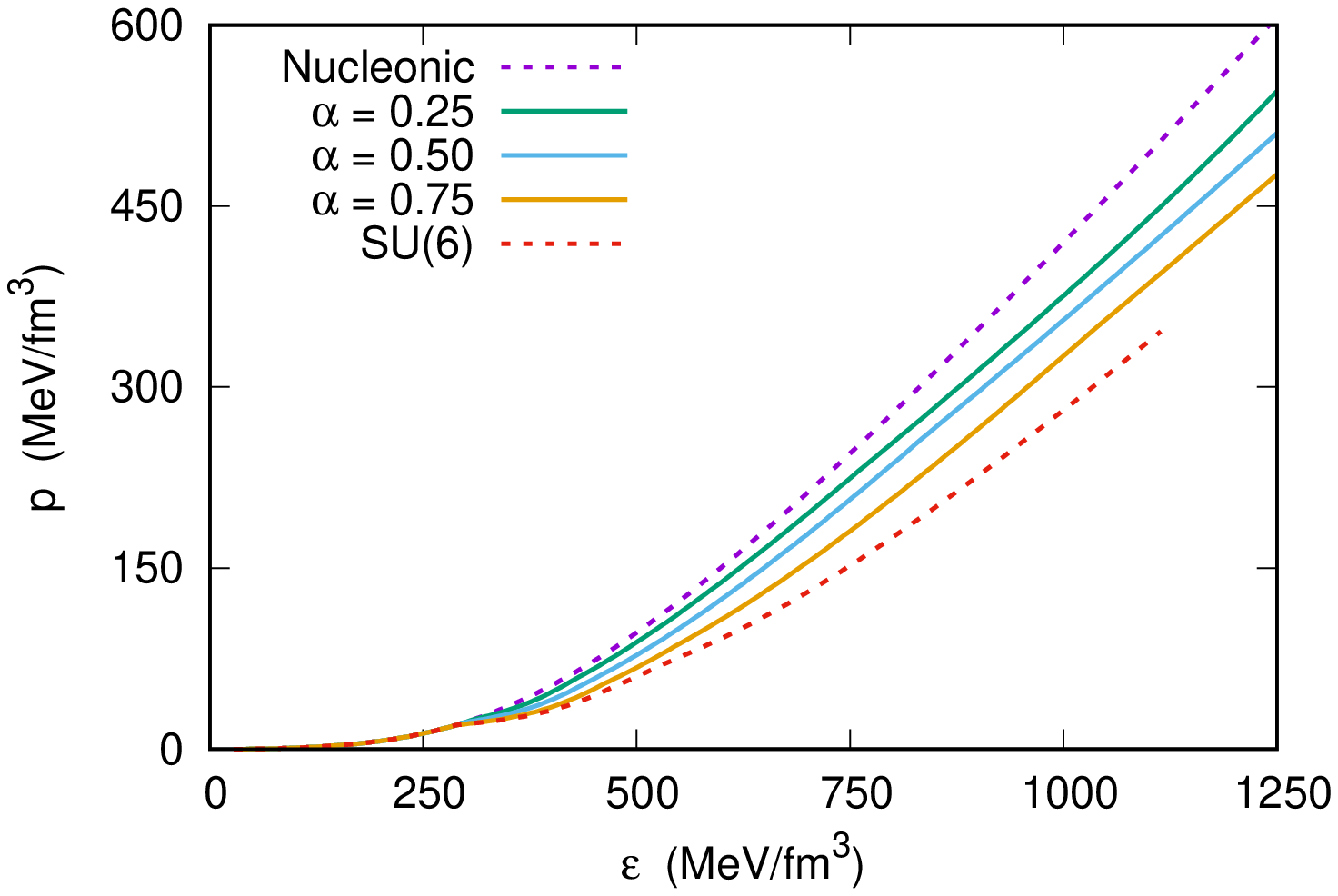} 

\includegraphics[angle=270,
width=0.46\textwidth]{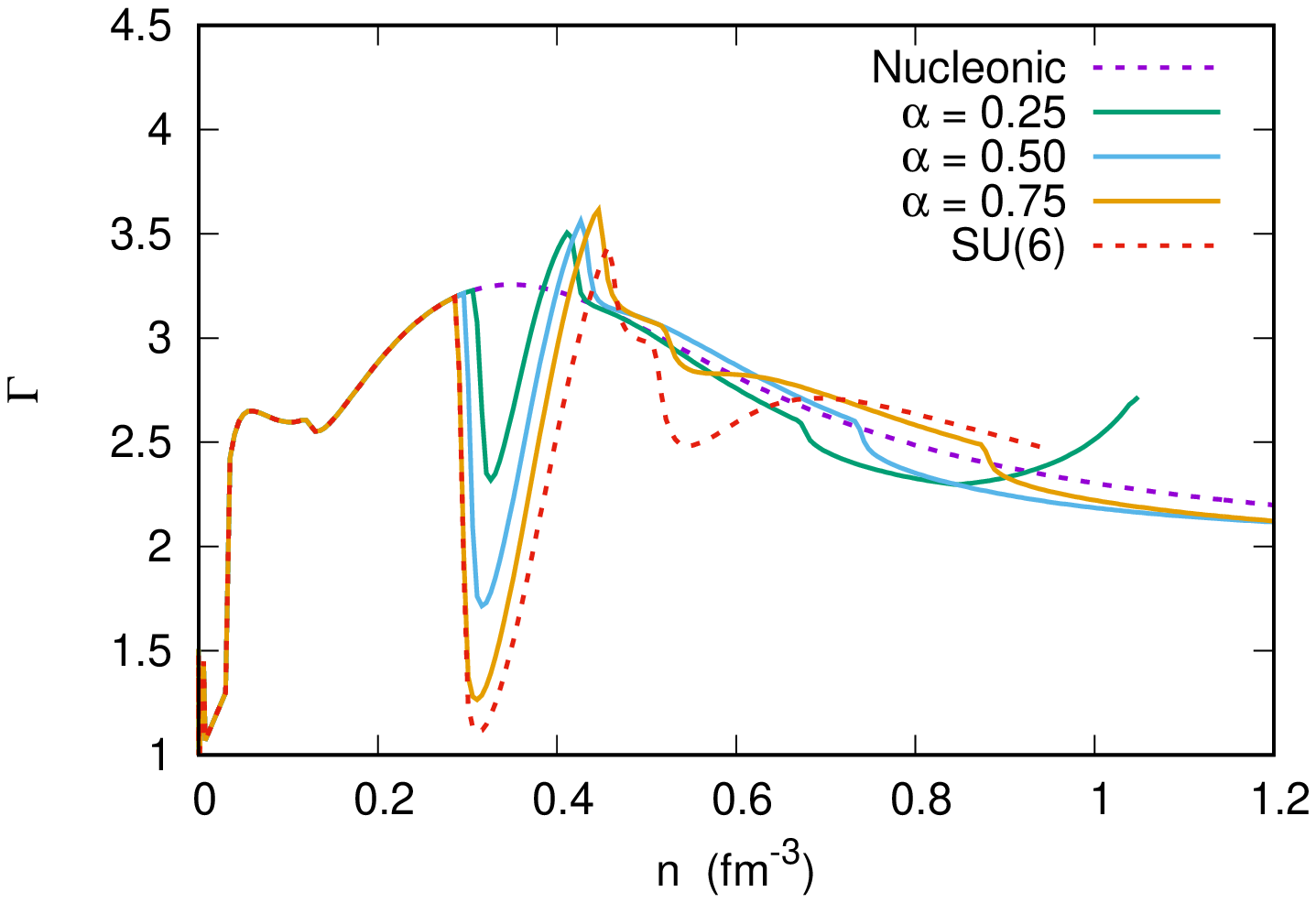}
\caption{(Color online)  Equation of state (top) and adiabatic index $\Gamma$ (bottom) for $\Delta$-admixed hyperonic star matter for different values of $\alpha_v$.  } \label{FL10}
\end{centering}
\end{figure}

We now display in Fig.~\ref{FL10} the EOS and the adiabatic index $\Gamma$ for the $\Delta$-admixed hyperonic star matter. As can be seen, the EOS are again distinguishable. The lower the value of $\alpha_v$ the stiffer the EOS, but unlike the pure $\Delta$-admixed matter case, we do not have any EOS stiffer than the pure nucleonic one, once the coupling constants with the hyperons are smaller. In relation to the adiabatic index, the behaviour of $\Gamma$ is even richer and more complex. We see a huge drop of $\Gamma$ due to the $\Delta^-$ threshold followed by a quick increase and a new drop due to the onset of the  $\Lambda^0$.

\begin{figure}[!t] 
\begin{centering}
 \includegraphics[angle=270,
width=0.46\textwidth]{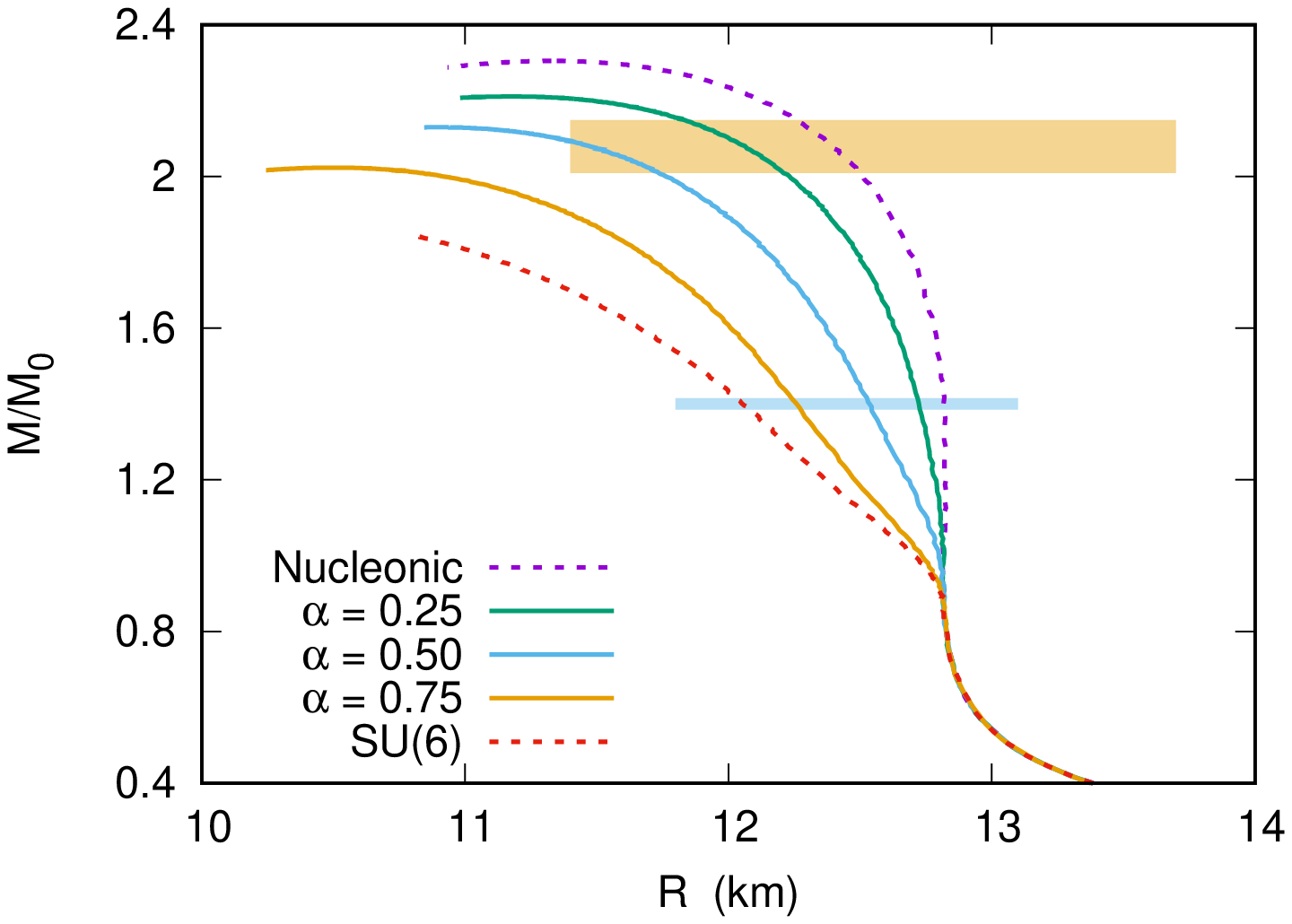} \\
\includegraphics[angle=270,
width=0.46\textwidth]{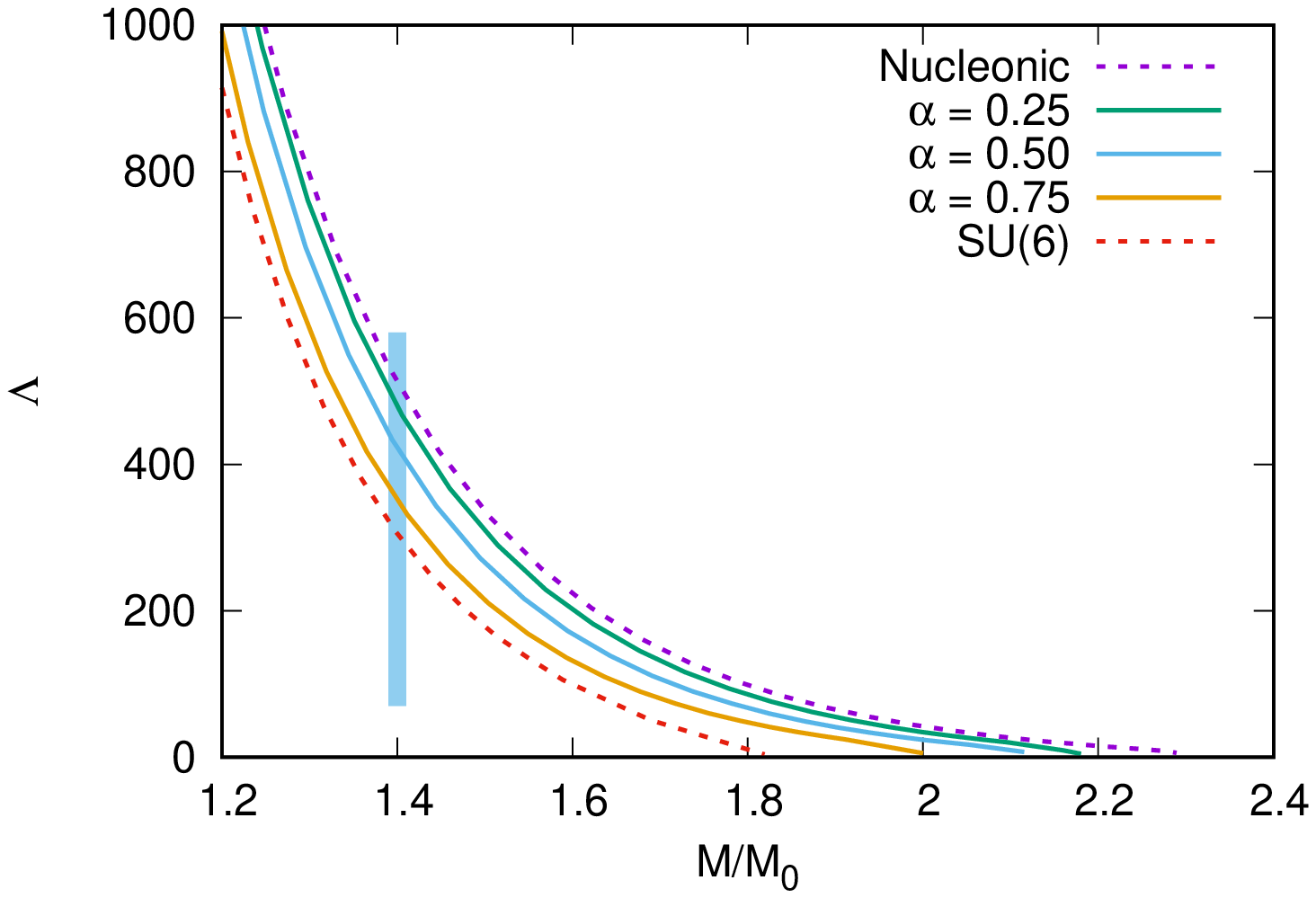}
\caption{(Color online) TOV solution (top) and dimensionless tidal parameter $\Lambda$ (bottom) for $\Delta$-admixed neutrons stars for different values of $\alpha_v$.  } \label{FL11}
\end{centering}
\end{figure}

Finally we display in Fig.~\ref{FL11} the TOV solution as well the dimensionless tidal parameter $\Lambda$. We see that we have an intermediate result between pure hyperonic neutron star and pure $\Delta$-admixed neutron star. In this case, the nucleonic neutron star has always a higher maximum mass.  Moreover, neither the SU(6) parameterization nor the $\alpha_v = 0.75$ can describe the PSR J0740+6620 pulsar, despite the fact that for $\alpha_v = 0.75$, a maximum mass of $M=2.02$  M$_\odot$ is reached, its radius is too low to cross the hatched area. 
For the SU(6) parameterization, a mass of $M=1.84$ M$_\odot$ is reached. However, we must emphasize that this does not represent the true maximum mass, as for this parameterization the numerical code stops before the condition $\partial M/ \partial \epsilon_c = 0$
is reached (see the discussion in Ref. \cite{nossoConstanca}). We want to note that the presence of hyperons and deltas produces the higher central density of the models under investigation, reaching $n=1.14$ fm$^{-3}$, which is over seven times the nuclear saturation density. Also, for $\alpha_v = 0.25$, a maximum mass in the range of the black widow pulsar is reached. It is also worth pointing out  that for $\alpha_v = 0.25$, the $\Delta$-admixed hyperonic neutron star reaches a  maximum mass higher than the pure hyperonic case. This counterintuitive behavior when deltas are included in the neutron  star matter, given the well-known hyperon puzzle reasoning, was first noticed and explained in Ref. \cite{nossoConstanca}. It occurs due to the increased repulsion between the baryons in the high density region of the star, because deltas have more repulsive couplings than the hyperons. Here, we presented an unified and model-independent formalism to determine the exotic particle couplings that strongly suggests that not only delta baryons must populate the neutron  star matter, but also that their presence makes the stars more massive.

Concerning the canonical star, we want to note that hyperons are present with a very low fraction only for the SU(6) parameterization. This reduces the radius for the canonical star to $R_{1.4}=12.05$ km, a small reduction when compared with the $R_{1.4}=12.08$ km for a pure $\Delta$-admixed neutron star. A reduction of the dimensionless tidal parameter from $\Lambda_{1.4}=318$ to $\Lambda_{1.4}=311$ is obtained. The macroscopic neutron star main results are summarized in Tab.~\ref{T5}.

\begin{table}[!t]
\begin{center}
\begin{tabular}{c|cccc|c}
&\multicolumn{4}{c|}{$\alpha_v$}&{ }\\
 $\alpha_v$ & SU(6) & 0.75 &  0.50 &  0.25  & Nucl. \\
\hline
 M$_{\rm max}$/M$_\odot$      & 1.84*  & 2.02   & 2.13 & 2.21 & 2.30   \\
 $n_c$ (fm$^{-3}$)        & 0.95* & 1.14  & 1.05 & 0.98 & 0.94  \\
 R  (km)       & 10.81* & 10.50  & 10.90 & 11.18 & 11.34  \\
 R$_{1.4}$  (km)          & 12.05 & 12.25  & 12.53 & 12.73 & 12.82  \\
$\Lambda_{1.4}$         & 311 & 360  & 428 & 489 & 516 \\
\hline
\end{tabular}
 \caption{$\Delta$-admixed hyperonic neutron star main properties for the maximum mass and canonical stars. The properties of the nucleonic star is included for comparison, and the * indicates that the true maximum mass was not reached.}\label{T5}
 \end{center}
 \end{table}
 
 We conclude that the $\Delta^-$ is by far the most important exotic particle that can be present in the neutron star core, once it appears at significantly low densities, and its fraction increases very rapidly 
 due to the very attractive potential, as well as its negative charge and spin 3/2. Moreover, based on the present study, it is the only particle able to influence even the canonical  star properties.

 \section{Conclusion}
 
 In this work we calculated the baryon-vector meson coupling constants for all members of the baryon octed and the baryon decuplet using the Clebsch-Gordan coefficients of the SU(3) group. Then, using a QHD model that virtually fulfills all the constraints at the saturation density, we fixed the coupling constants of the scalar $\sigma$ meson in order to reproduce the known potential depth, which allows an unified approach to hyperons and delta resonances coupling constants. We then apply the calculated coupling constants to study neutron star properties.  The main remarks are listed bellow.
 
 \begin{itemize}
 
 \item The vector meson coupling constants are
 fully model independent,  and is determined by fixing the free parameter $\alpha_v$.
  The results presented in Tab.~\ref{T2} can be applied to a diversity of problems that go beyond the one shown in the present paper.

 \item When we deal with hyperonic nuclear matter:

 \begin{itemize}
 
\item We see that the $\Lambda^0$ and $\Xi^-$ are always present, independent of the value of $\alpha_v$. These results corroborate the discussion presented in Ref.~\cite{lopesnpa,Dapo}, where it is stated that hyperons are inevitable.
 
 \item As expected from the hyperon puzzle reasoning, the hyperons soften the EOS. In a symmetry group approach, the higher the value of $\alpha_v$, the softer the EOS.
 
 \item The adiabatic index $\Gamma$ shows a huge drop due to the onset of the $\Lambda^0$, followed by a small one due to the $\Xi^-$ onset.
 
 \item With the exception of the SU(6) parameterization, we are able to explain the mass and the radius of the PSR J0740+6620 pulsar~\cite{NICER3}. For a low value of $\alpha_v$, even the mass range of the  black widow pulsar is reached.
 
 \item Within our model, hyperons are never present in a $M=1.4$  M$_\odot$ sta. Therefore, we always have $R_{1.4}$ = 12.82 km and $\Lambda_{1.4}$ = 516 if we consider a hyperonic neutron star matter composition.

\end{itemize}

 \item In the case of 
 $\Delta$-admixed nuclear matter: 

 \begin{itemize}
 
\item We see that the $\Delta^-$ is always present, in a density even lower than the $\Lambda^0$ onset in the hyperonic neutron star matter composition.  Also, the matter is
 strongly deleptonizated for lower values of $\alpha_v$. As in the case of hyperons, the $\Delta^-$ also seems inevitable, due to their strongly attractive potential, $U_\Delta = -98$ MeV and its negative charge.
 
 \item For higher values of $\alpha_v$, the $\Delta$-admixed nuclear matter is softer than the pure nucleonic one. But for lower values, this correlation is not so straightforward. In fact, the $\Delta$ matter can become even stiffer than the nucleonic one at higher densities.
 
 \item The adiabatic index $\Gamma$ for $\Delta$-admixed matter shows a much more complex behavior than the hyperonic one. We see a huge drop of $\Gamma$ due to the onset of the $\Delta^-$, which is followed by a quick increase. The $\Gamma$ at intermediate
 densities becomes even higher than for the pure nucleonic one, a behaviour not present in the hyperonic case.
 
 \item For all values of $\alpha_v$ we are able to reach a maximum mass above two solar masses; although for the SU(6) case the radii are too low and in disagreement with the inferred values for the PSR J0740+6620 pulsar~\cite{NICER3}. For low values of $\alpha_v$ the high mass of the PSR J0952-0607 pulsar is reached~\cite{Romani2022}.

\end{itemize}

  \item For the $\Delta$-admixed hyperonic nuclear matter:

\begin{itemize}  
  
\item  We see that the $\Delta^-$ and the $\Lambda^0$ are always present. Also, the $\Delta^-$ is always the first exotic particle to appear, followed by the $\Lambda^0$ that then becomes the more abundant particle in intermediate densities.
 
 \item As in the case previously analysed, the nucleonic EOS is always the stiffer one. Also the adiabatic index here is even richer and more complex than for pure $\Delta$-admixed matter. 

 \item With the SU(6) parameterization, the true maximum mass is not reached because the code stops converging still at low densities. This is the only case where this situation happens.
 For $\alpha_v$ lower or equal to 0.50, our results are able to describe the mass and radius of  the PSR J0740+6620 pulsar. For $\alpha_v$ = 0.25 even the very massive PSR J0952-0607 pulsar is also described.
 
 \item  Despite the fact that we use a very tight constraint for the radius of the canonical star, i.e., $R_{1.4}$ = 12.65 $\pm$ 0.45 km, with and uncertainty of only 5$\%$, all our models are able to satisfy such restriction. The same is true for the constraint of the dimensionless tidal parameter: 70 $ < \Lambda_{1.4} <$ 580.
 
 \item We conclude that, by far, the $\Delta^-$ is the most important exotic particle that can be present in the neutron star core. It is always present, for all parameterizations considered in the present work, and its onset happens at very low densities. Therefore the $\Delta^-$ affects the properties of the canonical 1.4 M$_\odot$. Also, its presence not always softens the EOS, but in some cases, the EOS can be even stiffer, increasing the maximum mass!
 
 \end{itemize}

  \end{itemize}

\begin{acknowledgments}
This  work is a part of the project INCT-FNA Proc. No. 464898/2014-5. D.P.M. is  partially supported by Conselho Nacional de Desenvolvimento Cient\'ifico e Tecnol\'ogico (CNPq/Brazil) under grant  301155/2017-8  and  K.D.M. under grant 150751/2022-2.
\end{acknowledgments}

\appendix 

\section{Yukawa couplings in SU(3) and SU(6) formalism}

In this appendix, we discuss in some detail the theory 
of the strong force, assuming that the Yukawa coupling of the QHD is invariant under the SU(3) flavor symmetry group.
In practice, this implies obtaining the SU(3) Clebsh-Gordan coefficients in order to keep the Yukawa Lagrangian density,
\begin{equation}
\mathcal{L}_{\rm Yukawa} = -g(\bar{\psi}_B\psi_B)M, \label{A1}
\end{equation}
as an unitary singlet. In Eq.~\eqref{A1}, $\psi_B$ is the Dirac field of the baryon $B$, $M$ is the field of an arbitrary meson and $\bar{\psi}_B$ is the complex conjugate of the field $\psi_B$. Each field is labeled by its eigenstates as
$|N, Y, I, I_3 \rangle$,  in conformance with the
usage of the SU(3) symmetry, where $N$ is dimension of the representation, $Y$ is the hypercharge, $I$ is the total isospin and $I_3$ is the isospin projection \cite{Swart1963,McNamee1964}. On the other hand, the complex conjugate of a state can be determined from
\begin{equation}
|N, Y, I, I_3 \rangle =  (-1)^{I_3 + Y/2} |\bar{N}, -Y, I, -I_3 \rangle . \label{A3}
\end{equation}
where $\bar{N}$ is the dimension of the representation of the complex conjugate.

Keeping the Yukawa coupling as an invariant implies the calculation of the SU(3) Clebsh-Gordan (CG) coefficients for the coupling $\bar{\psi}_B \otimes {\psi_B}\otimes M$.

\subsection{The baryon octet}

The use of the SU(3) formalism and the CG coefficients for the members of the baryon octet is well known in the literature ~\cite{Swart1963,McNamee1964,Pais,Lopes2013,lopesnpa,Lopes2020a,Weiss1,Weiss2,Dover1984,Tsu,MM1979,Stancu,Rijken,Lipkin}.
Nevertheless it is useful to retrieve the results, not only to explicitly show some subtleties, but also to help the reader who is unfamiliar with symmetry group techniques.

As we pointed out, the Lagrangian of  Eq.~\eqref{A1} must belong to the irreducible representation IR$\{1\}$, i.e, an
unitary singlet. Both the baryon octet ($\psi_B)$ and its complex conjugate ($\bar{\psi}_B$) belong to the IR$\{8\}$, $D(p,q)=D(1,1)$ \cite{Swart1963}. The vector meson nonet
belongs either to IR$\{8\}$ or to IR$\{1\}$. To preserve the unitary symmetry, the direct product $\bar{\psi}_B \otimes\psi_B$  must transform as IR$\{8\}$ when the meson eigenstate ($M$) belongs to IR$\{8\}$, and as IR$\{1\}$ when $M$
belongs to IR$\{1\}$.
By the use of the Speiser method~\cite{Swart1963}, the direct product $\{8\}$ $\otimes$  $\{8\}$  results in $\{27\}$ $\oplus$ $\{10\}$ $\oplus$ $\{\bar{10}\}$ $\oplus$ $\{8\}$ $\oplus$ $\{8'\}$ $\oplus$ $\{1\}$~\cite{Swart1963,Pais}.
Therefore, there are two ways to couple $\{8\}$ $\otimes$  $\{8\}$ to $\{8\}$; typically called antisymmetric  ($\{8\}$) and symmetric ($\{8'\}$) couplings~\cite{Stancu,Rijken}.

The Yukawa Lagrangian density can be rewritten as
\begin{equation}
\mathcal{L}_{\rm Yukawa} = -(gC + g'C' )(\bar{\psi}_B\psi_B)M, \label{A4}
\end{equation}
for the mesons belonging to IR$\{8\}$, and
\begin{equation}
\mathcal{L}_{\rm Yukawa} = -g_1(\bar{\psi}_B\psi_B)M, \label{A5}
\end{equation}
for the mesons belonging to IR$\{1\}$. 

The $g$ ($g'$) is the constant associated to the antisymmetric (symmetric) coupling, while the  $C$ ($C'$) is the SU(3) CG coefficients of the antisymmetric (symmetric) coupling.
The SU(3) CG coefficients can be calculated from the isoscalar factors, as discussed in Ref.~\cite{Swart1963}. Once its values are well known, we use the tables presented in Ref.~\cite{McNamee1964}. 

A crucial step is to realize that the CG coefficients presented in Ref.~\cite{McNamee1964} are related to the ($\psi_B \otimes \bar{\psi}_B$) direct product, instead of the ($\bar{\psi}_B \otimes \psi_B)$ presented in our Lagrangian. The CG in reversal order introduces a sign of $\pm 1$. As pointed in Tab. 1 from Ref.~\cite{McNamee1964}, the CG of the symmetric coupling ($C'$) gains a -1 sign. Without it, one cannot reproduce the results presented in Ref.~\cite{Swart1963,Dover1984,Weiss2,Lopes2013}.

Now, explicity, we have
\begin{align}
g_{NN\rho}={}& -\left( - \sqrt{\frac{3}{20}}g - \sqrt{\frac{1}{12}}g' \right) \times \sqrt{\frac{1}{8}}, \label{A6a} \\ 
g_{NN\omega_8}={}&- \left( \sqrt{\frac{1}{20}}g - \sqrt{\frac{1}{4}}g' \right) \times \sqrt{\frac{1}{8}},  
\end{align}
\begin{align}
g_{\Lambda\Lambda\rho}={}& 0,  \\
g_{\Lambda\Lambda\omega_8}={}&- \left(- \sqrt{\frac{1}{5}}g \right) \times - \sqrt{\frac{1}{8}}, 
\end{align}
\begin{align}
g_{\Sigma\Sigma_\rho}={}& - \left(- \sqrt{\frac{1}{3}}g' \right) \times  \sqrt{\frac{1}{8}},   \\ 
g_{\Sigma\Sigma\omega_8}={}&- \left(- \sqrt{\frac{1}{5}}g \right) \times  \sqrt{\frac{1}{8}}, 
\end{align}
\begin{align}
g_{\Xi\Xi\rho}={}& -\left( - \sqrt{\frac{3}{20}}g + \sqrt{\frac{1}{12}}g' \right) \times - \sqrt{\frac{1}{8}},  \\
g_{\Xi\Xi\omega_8}={}&- \left( - \sqrt{\frac{1}{20}}g - \sqrt{\frac{1}{4}}g' \right) \times - \sqrt{\frac{1}{8}}, \label{A6} 
\end{align}
and
\begin{equation}
  g_{NN\phi_1}=g_{\Lambda\Lambda\phi_1} = g_{\Sigma\Sigma\phi_1} = g_{\Xi\Xi\phi_1} = g_{1}.  
\end{equation}

Then, following Ref.~\cite{Swart1963}, we introduce the coupling constants:
\begin{equation}
g_8 =\frac{\sqrt{30}}{40} g + \frac{\sqrt{6}}{24 }g', \quad \mbox{and} \quad \alpha_v = \frac{\sqrt{6}}{24}\frac{g'}{g_8}, \label{A7}
\end{equation}
which allow us to rewrite the coupling constants of the baryons with the vector mesons (\ref{A6a}--\ref{A6}) in its more usual way~\cite{Swart1963,Dover1984,Lopes2013},
\begin{align}
g_{NN\rho} ={}& g_8, \\ g_{\Sigma\Sigma\rho}={}&2g_8\alpha_v, \\ g_{\Xi\Xi\rho}={}&-g_8(1 - 2\alpha_v),  \\
g_{\Lambda\Lambda\rho}={}&0,
\end{align}
and
\begin{align}
g_{NN\omega_8} ={}&\frac{1}{3}g_8\sqrt{3}(4\alpha_v -1),  \\
g_{\Lambda\Lambda\omega_8}={}&- \frac{2}{3}g_8\sqrt{3}(1 - \alpha_v),\\ g_{\Sigma\Sigma\omega_8}={}& \frac{2}{3}g_8\sqrt{3}(1 - \alpha_v),  \\
g_{\Xi\Xi\omega_8}={}&- \frac{1}{3}g_8\sqrt{3}(1 + 2\alpha_v).\label{A8}
\end{align}

In nature, nevertheless, the physical realisation of the mesons are the $\omega$ and $\phi$ meson, that are a mixture of the theoretical $\omega_8$ and $\phi_1$ states \cite{Dover1984},
\begin{align}
   \omega ={}& \cos\theta_v |\phi_1 \rangle + \sin\theta_v |\omega_8 \rangle  \label{A9a}   \\
  \phi ={}&  - \sin\theta_v |\phi_1 \rangle + \cos\theta_v |\omega_8 \rangle . \label{A9}  
\end{align}

Therefore, the coupling constants of the baryon octet with the $\omega$  meson now read
\begin{align}
g_{NN\omega}={}&g_1 \cos\theta_v + g_8\sin\theta_v\frac{1}{3}\sqrt{3}(4\alpha_v - 1), \label{A10a} \\
g_{\Lambda\Lambda\omega}={}&g_1 \cos\theta_v + g_8\sin\theta_v\frac{2}{3}\sqrt{3}(1 - \alpha_v),  \\
g_{\Sigma\Sigma\omega}={}&g_1 \cos\theta_v - g_8\sin\theta_v\frac{2}{3}\sqrt{3}(1 - \alpha_v),  \\
g_{\Xi\Xi\omega}={}&g_1 \cos\theta_v -  g_8\sin\theta_v\frac{1}{3}\sqrt{3}(1 + 2\alpha_v). \label{A10}
\end{align}
The results for the $\phi$ meson follow directly by 
 replacing $\cos\theta_v$ $\rightarrow$ $-\sin\theta_v$ and 
$\sin\theta_v$ $\rightarrow$ $\cos\theta_v$ in  the equations above \cite{Dover1984}.

Within the flavor SU(3) symmetry, we have in principle three free parameters ($\alpha_v$, the ratio $z=g_8/g_1$, and the mixing angle $\theta_v$). The mixing angle $\theta_v$ is related to the nature of the $\omega$ and $\phi$ mesons. When we assume SU(6) symmetry ($\theta_v=35.264$) we have the quark contents  $\omega = (\bar{u}u + \bar{d}d)/\sqrt{2}$ and $\phi$=$\bar{s}s$. This is called ideal mixing~\cite{Dover1984}. The $z$ and $\alpha_v$ parameters are related to the relative  strength of the coupling, and to the nature of the coupling itself. $\alpha_v$ = $F/(F+D)$  is a weight factor for the contributions of the
symmetric $D$ (corresponding to  $\{8'\}$) and the antisymmetric $F$ (corresponding to $\{8\}$) couplings relative to each other~\cite{Weiss2,MM1979,Rijken}. The ratio $z$ is the relative strength of the coupling of the  baryons with the meson octet over the singlet one. 

All three free parameters can be fixed by imposing that the Yukawa Lagrangian density is not only 
invariant under the flavor SU(3) symmetry group but also to the spin SU(2) symmetry group, creating  the so-called SU(6) hybrid group $SU(6) \supset SU(3) \otimes SU(2)$, which implies~\cite{Dover1984,Stancu,Pais,Lipkin,Rijken} that
\begin{equation}
z = \frac{1}{\sqrt{6}}, \quad \theta_v = 35.264, \quad \mbox{and} \quad \alpha_v = 1.00.\label{A11}
\end{equation}
In this case
we obtain that the $\phi$ meson does not couple to the nucleon ($g_{NN\phi} = 0$); the $\omega$ meson couples to the hyperchage and the $\rho$ meson couples to the isospin.
Such coupling nature was originally proposed by Sakurai~\cite{Sakurai}.

In this work, in order to study the effect of different coupling constants, we assume that the SU(3) flavor symmetry is exact but the hybrid SU(6) symmetry can be partially broken. In other words, all baryon-vector meson coupling constants obey Eq.~(\label{A10a}--\ref{A10}) and the values of $z$ and $\theta_v$ given in  Eq.~\eqref{A11} are kept fixed 
in agreement with the SU(6) symmetry, but $\alpha_v$ is left as a free parameter.

\subsection{The baryon decuplet}

Once we set the grounds and pave our knowledge to construct an Yukawa coupling which is invariant under the flavor SU(3) symmetry group for the baryon octet, we give a step further and  impose that the Yukawa coupling  is also invariant 
for the baryon decuplet. As one could correctly deduce, this implies calculating the CG coefficients for the baryon decuplet.

The baryon decuplet belongs to IR$\{10\}$, $D(p,q) =(3,0)$, while its complex conjugate belongs to IR$\{10^{*}\}$, $D(p,q) = D(0,3)$ ~\cite{Swart1963,Stancu}. The vector meson nonet belongs either to IR$\{8\}$ or to IR$\{1\}$. Therefore, to preserve the unitary symmetry, as in the case of the baryon octet, the direct product must transform as IR$\{8\}$ when the meson eigenstate ($M$) belongs to  IR$\{8\}$ and  as IR$\{1\}$ when $M$ belongs to  IR$\{1\}$.

Applying the Speiser method, the direct product $\{10\}$ $\otimes$  $\{10^{*}\}$  results in $\{64\}$ $\oplus$ $\{27\}$ $\oplus$  $\{8\}$ $\oplus$ $\{1\}$~\cite{McNamee1964}. Here, unlike the octet case, we have only one coupling resulting in a IR$\{8\}$, the antisymmetric one.

The Yukawa Lagrangian density can be rewritten as
\begin{equation}
\mathcal{L}_{\rm Yukawa} = -(gC)(\bar{\psi}_B\psi_B)M, \label{A12}
\end{equation}
for the mesons belonging to IR$\{8\}$, and
\begin{equation}
\mathcal{L}_{\rm Yukawa} = -g_1(\bar{\psi}_B\psi_B)M, \label{A13}
\end{equation}
for the mesons belonging to IR$\{1\}$, where $C$ are the SU(3) CG coefficients. The CG are listed in Ref.~\cite{McNamee1964}. It is also worth pointing out that the direct product that results in a IR$\{1\}$ must gain a -1 sign, as displayed in Tab. 1 from R.~\cite{McNamee1964}. 

In the case of the baryon octet, we have some doublet isospin, i.e, particles with the same eigenstates except by the sign of the isospin projection.  For instance, protons and neutrons form the nucleon  doublet: $N = |8, 1, 1/2, \pm 1/2 \rangle$; $\Xi^0$ and $\Xi^{-}$ correspond to $\Xi = |8, -1, 1/2, \pm 1/2 \rangle$. Here, it is useful to divide the four $\Delta$'s into two isospin doublets. We therefore define: ($\Delta^0,\Delta^+$) = $\Delta = |10, 1, 3/2, \pm 1/2 \rangle$, and ($\Delta^-, \Delta^{++}$) = $\Delta^* = |10, 1, 3/2, \pm 3/2 \rangle$. Within these definitions we have
\begin{align}
g_{\Delta^\ast\Delta^\ast\rho}={}&-\left(  \sqrt{\frac{3}{10}}g  \right) \times - \sqrt{\frac{1}{10}},  \\ 
g_{\Delta^\ast\Delta^\ast\omega_8}={}&- \left( \sqrt{\frac{1}{10}}g  \right) \times -\sqrt{\frac{1}{10}},   \\ ´
g_{\Delta\Delta\rho}={}&-\left( -\sqrt{\frac{1}{30}}g\right) \times \sqrt{\frac{1}{10}},  \\ 
g_{\Delta\Delta\omega_8}={}&- \left(- \sqrt{\frac{1}{10}}g  \right) \times \sqrt{\frac{1}{10}},   
\end{align}
\begin{align}
g_{\Sigma^{\ast }\Sigma^{\ast }_\rho}={}&- \left(- \sqrt{\frac{2}{15}}g' \right) \times  \sqrt{\frac{1}{10}},   \\ 
g_{\Sigma^{\ast }\Sigma^{\ast }\omega_8}={}&0,  
\end{align}
\begin{align}
g_{\Xi^\ast \Xi^\ast \rho}={}&-\left( \sqrt{\frac{1}{30}}g \right) \times - \sqrt{\frac{1}{10}},  \\
g_{\Xi^\ast \Xi^\ast \omega_8}={}&- \left( - \sqrt{\frac{1}{10}}g  \right) \times - \sqrt{\frac{1}{10}}, 
\end{align}
\begin{align}
g_{\Omega\Omega\rho}={}&0,  \\
g_{\Omega\Omega\omega_8}={}&- \left(  \sqrt{\frac{2}{5}}g \right) \times  \sqrt{\frac{1}{10}}, \label{A14} 
\end{align}
and
\begin{align}
  g_{1}={}&g_{\Delta\Delta\phi_1} = g_{\Delta^\ast \Delta^\ast \phi_1} \nonumber\\ ={}&g_{\Sigma^\ast \Sigma^\ast \phi_1}   =  g_{\Xi^\ast \Xi^\ast \phi_1} = g_{\Omega\Omega\phi_1}.  
\end{align}

Now, in analogy with Eq.~\eqref{A7} related to the baryon octet, we introduce a new coupling constant to the baryon decuplet
\begin{equation}
g_{10} = 10\sqrt{3}g. \label{A15}
\end{equation}

As the real $\phi$ and $\omega$ are are not the theoretical $\omega_8$ and $\phi_1$ but a mixture of them, as put in the Eqs.~ (\ref{A9a}--\ref{A9}), we can rewrite the coupling constants for the baryon decuplet as
\begin{align}
g_{\Delta^*\Delta^*\rho}={}&3g_{10}, \\ g_{\Delta\Delta\rho}={}&g_{10}, \\ g_{\Sigma^*\Sigma^*\rho}={}&2g_{10}, \\
g_{\Xi^*\Xi^*\rho}={}&g_{10}, \\ g_{\Omega\Omega\rho}={}&0,
\end{align}
and
\begin{align}
g_{\Delta^*\Delta^*\omega}={}&g_{\Delta\Delta\omega}=g_1\cos\theta_v + g_{10}\sqrt{3}\sin\theta_v ,\\
g_{\Sigma^{*}\Sigma^{*}\omega}={}&g_1\cos\theta_v , \\
g_{\Xi^*\Xi^*\omega}={}&g_1\cos\theta_v - g_{10}\sqrt{3}\sin\theta_v ,\\ 
g_{\Omega\Omega\omega}={}&g_1\cos\theta_v - g_{10}\sqrt{12}\sin\theta_v , \label{A16}
\end{align}
and again, the results for the $\phi$ meson are obtained, by  replacing $\cos\theta_v \rightarrow -\sin\theta_v$ and  $\sin\theta_v\rightarrow \cos\theta_v$.

As can be seem, for the baryon decuplet we have only two instead of three free parameters. As a consequence, the coupling for the $\rho$ meson is already determined.
Now we employ SU(6) symmetry, and assume an ideal mixing angle ($\theta_v=35.264$). We also determine $z$ in such a way that $g_{\Delta^*\Delta^*\phi} = g_{\Delta\Delta\phi}$ = 0, which implies $z = 1/\sqrt{6}$, exactly as in the octet case.

\subsection{The inter-multiplet ratio}

If we want to describe  on the same footing, a strongly interacting matter composed of both the baryon octet and decuplet members, then the ratios of the type $g_{\Delta\Delta\omega}/g_{NN\omega}$ are crucial.
For simplicity, but without any loss of generality, we start by assuming a full SU(6) parameterization, i.e, $\alpha_v$ = 1.00. In this case we have
\begin{equation}
\frac{g_{\Delta\Delta\omega}}{g_{NN\omega}} = \chi_{\Delta\omega} =  \frac{g_{10}}{g_8} . \label{A17}
\end{equation}

From a pure, abstract, symmetry group theory arguments, the baryon octet and decuplet are unrelated multiplets of the SU(3) group. Therefore, in principle, $\chi_{\Delta\omega}$ can assume any value. 
However, we known that the nucleon and the $\Delta$ are not pure abstract entities. Indeed, they can be faced as the same particle, being the $\Delta$ only an excited state of the nucleon. We also know that in accordance with the SU(6) symmetry, for the members of the same multiplet the $\omega$ meson couples to the hypercharge. This is also in agreement with the Sakurai's theory of the strong interaction~\cite{Sakurai}.
Therefore, it is reasonable to assume that the nature of the coupling with the $\omega$ meson holds even in the inter-multipet case. So, we impose
\begin{equation}
\frac{g_{10}}{g_8} = 1 . \label{A18}
\end{equation}
Once we fixed the the ratio $g_{10}/g_8$, for a given value of $\alpha_v$ (assuming that $\theta_v$ and $z$ obey Eq.~\ref{A11}) every single relative coupling of the baryon octet and decuplet with the vector mesons are fully  determined.

\end{document}